\documentclass[lettersize,journal]{IEEEtran}
\usepackage{amsmath,amsfonts}
\usepackage{algorithmic}
\usepackage{array}
\usepackage[caption=false,font=normalsize,labelfont=sf,textfont=sf]{subfig}
\usepackage{textcomp}
\usepackage{stfloats}
\usepackage{url}
\usepackage{verbatim}
\usepackage{graphicx}
\usepackage{siunitx}
\usepackage{booktabs}
\def\BibTeX{{\rm B\kern-.05em{\sc i\kern-.025em b}\kern-.08em
    T\kern-.1667em\lower.7ex\hbox{E}\kern-.125emX}}
\usepackage{balance}

\usepackage{pifont}
\usepackage{multirow}
\usepackage{acro}
\usepackage{hyperref}
\usepackage{cite}
\usepackage{soul}
\usepackage{microtype}
\usepackage{xspace}
\usepackage{textcomp}
\usepackage{tabularx}
\DeclareAcronym{davenet}{
	short = DAVEnet ,
	long  = deep audio-visual embedding network,
	alt = Deep Audio-Visual Embedding Network
}
\DeclareAcronym{resdavenet}{
	short = ResDAVEnet,
	long  = residual deep audio-visual embedding network,
	alt = Residual Deep Audio-Visual Embedding network
}
\DeclareAcronym{AE}{
	short = AE ,
	long  = autoencoder
}
\DeclareAcronym{CAE}{
	short = CAE,
	long  = correspondence autoencoder,
	alt = Correspondence Autoencoder
}
\DeclareAcronym{MCAE}{
	short = MCAE,
	long  = multimodal correspondence autoencoder,
	alt = Multimodal Correspondence Autoencoder
}
\DeclareAcronym{MTriplet}{
	short = MTriplet,
	long  = multimodal triplet network,
	alt = Multimodal Triplet Network
}
\DeclareAcronym{FFNN}{
	short = FFNN,
	long  = feedforward neural network,
	alt = Feedforward Neural Networks
}
\DeclareAcronym{CNN}{
	short = CNN,
	long  = convolutional neural network,
	alt = Convolutional Neural Networks
}
\DeclareAcronym{RNN}{
	short = RNN,
	long  = recurrent neural network,
	alt = Recurrent Neural Networks
}
\DeclareAcronym{VQ}{
	short = VQ,
	long  = vector-quantisation
}
\DeclareAcronym{mfcc}{
	short = MFCC,
	long  = mel-frequency cepstral coefficient
}
\DeclareAcronym{CPC}{
	short = CPC,
	long  = contrastive predictive coding,
	alt = Contrastive Predictive Coding
}
\DeclareAcronym{multdavenet}{
	short = MultDAVEnet,
	long  = multilingual deep audio-visual embedding network,
	alt = Multilingual Deep Audio-Visual Embedding Network
}
\DeclareAcronym{accmultdavenet}{
	short = AcstMultDAVEnet,
	long  = acoustic multilingual deep audio-visual embedding network,
	alt = Acoustic Multilingual Deep Audio-Visual Embedding Network
}
\DeclareAcronym{DNN}{
	short = DNN,
	long  = deep neural network
}
\DeclareAcronym{ASR}{
	short = ASR,
	long  = automatic speech recognition
}
\DeclareAcronym{AWE}{
	short = AWE,
	long  = acoustic word embedding,
	alt = Acoustic Word Embedding
}
\DeclareAcronym{AGWE}{
	short = AGWE,
	long  = acoustically grounded word embedding,
	alt = Acoustically Grounded Word Embedding
}
\DeclareAcronym{ASE}{
	short = ASE,
	long  = acoustic span embedding,
	alt = Acoustic Span Embedding
}
\DeclareAcronym{QbE}{
	short = QbE,
	long  = query-by-example,
	alt  = Query-by-Example
}
\DeclareAcronym{STD}{
	short = STD,
	long  = spoken term detection
}
\DeclareAcronym{UTD}{
	short = UTD, 
	long = long short-term memory
}
\DeclareAcronym{SGD}{
	short = SGD, 
	long = stochastic gradient decent,
	alt = Stochastic Gradient Decent
}
\DeclareAcronym{bow}{
	short = BoW, 
	long = bag-of-words,
	alt = Bag-of-Words
}
\DeclareAcronym{mse}{
	short = MSE, 
	long = mean squared error,
	alt = Mean Squared Error
}
\DeclareAcronym{vpkl}{
	short = VPKL, 
	long = visually prompted keyword localisation,
	alt = Visually Prompted Keyword Localisation
}

\makeatletter
\def\blfootnote{\gdef\@thefnmark{}\@footnotetext}
\makeatother

\newcommand{\model}{{\small \scshape MattNet}}
\newcommand{\smallmodel}{{\scriptsize \scshape MattNet}}
\newcommand{\qbert}{{\small QbERT}}

\newcommand{\boy}{\d{o}m\d{o}k\`{u}nrin}
\newcommand{\dogs}{\`{a}w\d{o}n aj\'{a}}
\newcommand{\grass}{kor\'{i}ko}
\newcommand{\rock}{\`{a}p\'{a}ta}
\newcommand{\water}{omi}
\newcommand{\yoruba}{Yor\`{u}b\'{a}}

\newcommand{\captionsep}{\vspace*{-5pt}}

\usepackage[prependcaption,textsize=scriptsize]{todonotes}
\definecolor{mycolor}{HTML}{008000}%
\definecolor{indiagreen}{HTML}{138808}%
\definecolor{papaya}{HTML}{EE892F}%
\definecolor{mygreen}{HTML}{008000}%
\definecolor{mypurple}{HTML}{9966CC}%
\definecolor{myblue}{HTML}{5D8AA8}

\usepackage{combelow}

\begin{document}
\title{Visually grounded few-shot word learning in low-resource settings}%
\author{Leanne Nortje, Dan Onea\cb{t}\u{a}, Herman Kamper
\thanks{Leanne Nortje and Herman Kamper are with the Department of Electrical and Electronic Engineering, Stellenbosch University, Stellenbosch, 7600, South Africa (email: \href{mailto:nortjeleanne@gmail.com}{nortjeleanne@gmail.com}; \href{mailto:kamperh@sun.ac.za}{kamperh@sun.ac.za}).} %
\thanks{Dan Onea\cb{t}\u{a} is with the University Politehnica of Bucharest,
Bucharest, RO-060042, Romania (e-mail: \href{mailto:dan.oneata@gmail.com}{dan.oneata@gmail.com}).} %
\thanks{%
Google DeepMind funded Leanne Nortje.
Dan Oneață was supported by European Union’s HORIZON-CL4-2021-HUMAN-01 research and innovation program under grant agreement No. 101070190 AI4Trust.
We would like to thank Tyler Miller and David Harwath for helping with the few-shot retrieval comparisons, and Benjamin van Niekerk for helping with \qbert.}}

\markboth{Submitted for review, 2023}%
{How to Use the IEEEtran \LaTeX \ Templates}

\maketitle

\begin{abstract}
    We propose a visually grounded speech model that learns new words and their visual depictions from just a few word-image example pairs. 
    Given a set of test images and a spoken query, we ask the model which image depicts the query word.
    Previous work has simplified this few-shot learning problem by either using an artificial setting with digit word-image pairs or by using a large number of examples per class.
    Moreover, all previous studies were performed using English speech-image data.
    We propose an approach that can work on natural word-image pairs but with less examples, i.e.\ fewer shots, and then illustrate how this  approach can be applied for multimodal few-shot learning in a real low-resource language, \yoruba.
    Our approach involves using the given word-image example pairs to mine new unsupervised word-image training pairs from large collections of unlabelled speech and images.
    Additionally, we use a word-to-image attention mechanism to determine word-image similarity. 
    With this new model, we achieve better performance with fewer shots than %
    previous approaches on an existing English benchmark.
    Many of the model's mistakes are due to confusion between visual concepts co-occurring in similar contexts.
    The experiments on \yoruba\ show the benefit of transferring knowledge from a multimodal model trained on a larger set of English speech-image data.%
    \footnote{Code and models: \href{https://sites.google.com/view/few-shotwordacquisition/home}{Project webpage}.}
\end{abstract}


\section{Introduction}
\label{sec:intro}

Speech recognition for low-resource languages faces a major obstacle: it requires large amounts of transcribed data for development~\cite{besacier_automatic_2014}.
In some extreme cases, it might even be impossible to get any labelled data, e.g.\ when dealing with an unwritten language. 
This is in stark contrast to infants that learn words without access to any transcriptions
\cite{biederman_recognition-by-components:_1987, miller_how_1987, gomez_infant_2000, lake_one-shot_2014, rasanen_joint_2015}.
This is one motivation for
recent studies into multimodal few-shot learning~\cite{eloff_multimodal_2019, nortje_unsupervised_2020, nortje_direct_2021}: %
learning new concepts from a few examples, where each example consists of instances of the same concept but from different modalities. 
E.g., imagine a robot seeing a picture of a \textit{zebra}, \textit{kite} and \textit{sheep}
while also hearing the spoken word for each concept. 
After seeing this small set of examples (called a support set) 
the robot is prompted to identify which image in an unseen set corresponds to the word ``zebra''.

Building off of a growing number of studies in visually grounded speech modelling~\cite{harwath_jointly_2018, kamper_semantic_2019-1, olaleye_attention-based_2021, chrupala_visually_2022, merkx_modelling_2022, peng_syllable_2023, berry_m-speechclip_2023}, we consider this multimodal problem of learning the spoken form of a word and its visual depiction from only a few paired word-image examples.
Multimodal few-shot speech-image learning was first introduced in~\cite{eloff_multimodal_2019} and then extended in~\cite{nortje_unsupervised_2020} and~\cite{nortje_direct_2021}.
But these studies were performed in an artificial setting where spoken isolated digits were paired with MNIST images of digits.
This shortcoming was recently addressed by Miller and Harwath~\cite{miller_tyler_exploring_2022}, who considered multimodal few-shot learning on isolated words paired with natural images.
Their specific focus was on learning a new concept while not forgetting previously learned concepts, i.e.\ dealing with the problem of catastrophic forgetting.  
(We do not explicitly focus on the catastrophic forgetting problem here, although we do evaluate using the same setup as~\cite{miller_tyler_exploring_2022}.)
While their methods performed well in a few-shot retrieval task with five classes, they required a relatively large number of samples per class, i.e. many ``shots''.
Our first overarching aim is to do visually grounded multimodal few-shot learning on  natural images with fewer shots.
All previous studies also performed experiments using English speech-image data. Our second goal, therefore, is to  present a few-shot evaluation on a real low-resource language.

Our new multimodal few-shot approach combines two core ideas.
Firstly, we use the support set to ``mine'' new noisy word-image pairs from unlabelled speech and image collections.
Concretely, each spoken word example in the support set is compared to each utterance in an unlabelled speech corpus; 
we use a new query-by-example approach %
to identify segments in the search utterances that match the word in the support set.
We follow a similar approach for mining additional images from the few-shot classes by using cosine distance between pretrained image
embeddings.
The mined words and images are then paired up, thereby artificially increasing the size of our support set in an unsupervised way.
This %
mining scheme is very similar to that followed in~\cite{nortje_direct_2021}, where it was used on digit image-speech data with simpler within-modality comparisons.
Secondly, our new approach is based on a model with a word-to-image attention mechanism.
This multimodal attention network (\model), takes a single word embedding and calculates its correspondence to each pixel embedding to learn how the word is depicted within an image.
This is similar to the vision attention part of the model from~\cite{nortje_towards_2022}, where the goal was to localise visual keywords in speech (not in a few-shot setting).

Using the English SpokenCOCO speech-image dataset~\cite{hsu_text-free_2021}, two evaluation settings are considered.
We first evaluate our approach on the few-shot retrieval task also used in~\cite{miller_tyler_exploring_2022}.
We show that \model\ achieves higher retrieval scores for fewer shots than \cite{miller_tyler_exploring_2022}'s models.
Secondly, we evaluate our approach in a more conventional few-shot classification task where it only needs to correctly distinguish between classes seen in the support set. 
We show that we can achieve five-way accuracies higher than 80\% with as little as five shots.

Still using the English SpokenCOCO models, we then perform an exhaustive analysis to understand the parts of our approach that are most essential, the characteristics of mistakes, and how performance differs across keywords. This  includes qualitative results showing that the few-shot model can localise objects in images given a spoken word query.

To address our second overarching goal, we finally do multimodal few-shot learning using a \yoruba\ speech-image dataset~\cite{olaleye_yfacc_2023}, illustrating for the first time that these few-shot approaches can be applied to a real low-resource language.

This work is an extension of the conference paper~\cite{nortje_visually_2023}, where the main results on the SpokenCOCO benchmark were presented (a slightly improved model is used here).
The current paper extends this work with a thorough  analysis, ablation experiments, and the application of the model to a low-resource language. To summarise, we make the following contributions: (1) we introduce a new mining scheme operating on natural images and speech, (2) we introduce a new attention-based model for multimodal few-shot learning, (3) we give a thorough analysis of the proposed approach, and (4) apply the approach in a new low-resource multimodal few-shot learning benchmark.

\section{Visually grounded few-shot learning and evaluation}
\label{sec:task}

We train a model on a few spoken word-image examples.
The set of $K$ examples per class is called the support set $\mathcal{S}$.
Each pair in $\mathcal{S}$ consists of an isolated spoken word $\boldsymbol{\mathrm{a}}_{j}$ and a corresponding image $\boldsymbol{\mathrm{v}}_{j}$.
For the \textit{one-shot} case shown in the top part of Fig.~\ref{fig:task}, $\mathcal{S}$ consists of one word-image example pair for each of the $L$ classes. 
For the $L$-way $K$-shot task, the support set $\mathcal{S}=\{\boldsymbol{\mathrm{a}}_{j}, \boldsymbol{\mathrm{v}}_{j}\}_{\scriptscriptstyle j=1}^{\scriptscriptstyle L\times K}$ contains $K$ word-image example pairs for each of the $L$ classes.  
In this work, we use a few-shot model for two tasks, as we describe next.

\subsection{Visually grounded few-shot word classification}
\label{subsec:classification_task}
\begin{figure}[!t]
	\centering
	\includegraphics[width=0.9\linewidth]{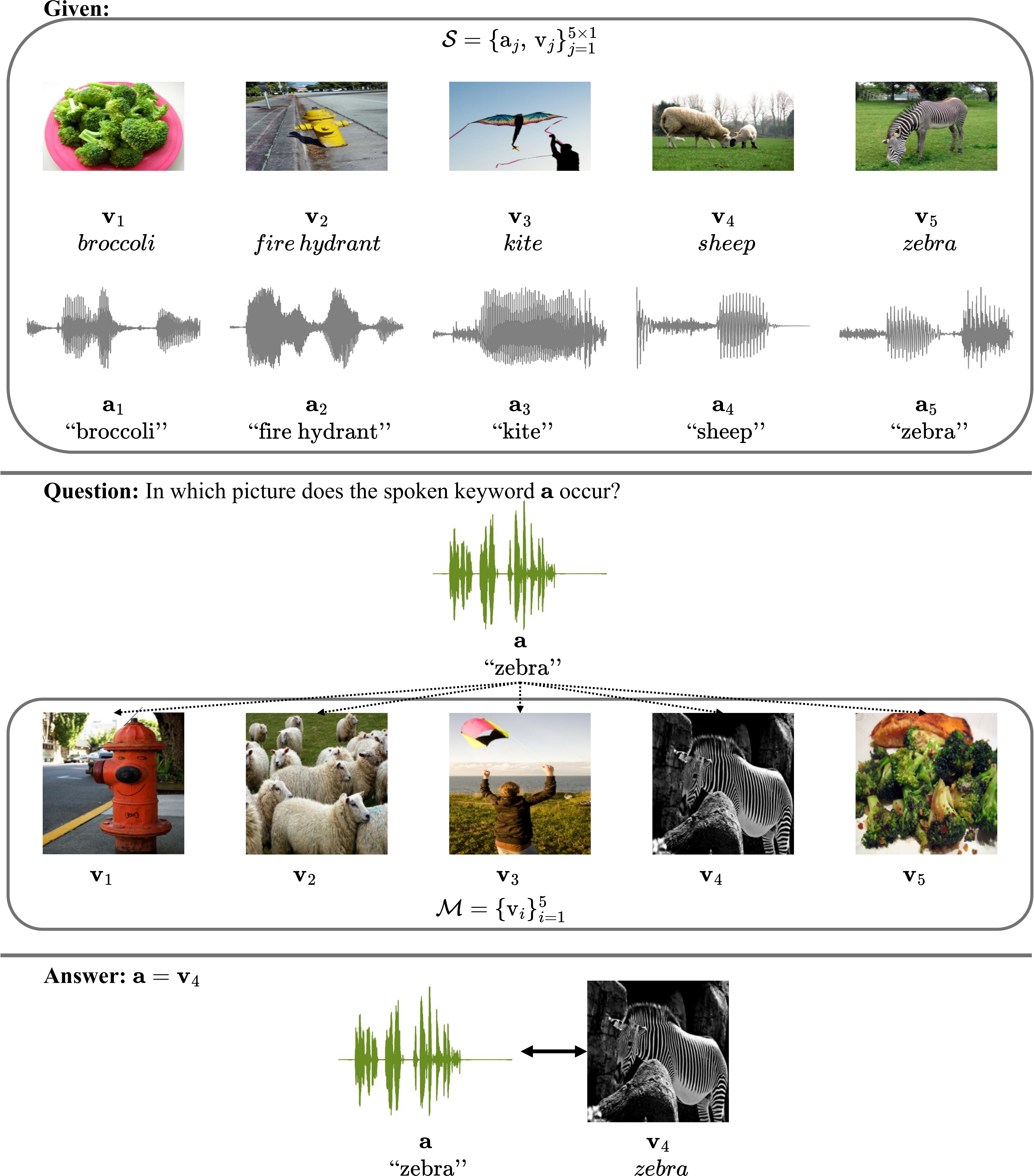}
	\caption{Given the few examples in the support set $\mathcal{S}$, the multimodal few-shot classification task is to e.g.\ identify the image depicting the word ``zebra'' from a set of unseen images.}
	\label{fig:task}
\end{figure}

In this task, illustrated in the middle and bottom
of Fig.~\ref{fig:task}, we are given an unseen isolated spoken word query $\boldsymbol{\mathrm{a}}$ and prompted to identify the corresponding image in a matching set  $\mathcal{M}=\{\boldsymbol{\mathrm{v}}_{i}\}_{\scriptscriptstyle i=1}^{\scriptscriptstyle L}$ of unseen test images.
$\mathcal{M}$ contains one image depicting each of the $L$ classes.
Neither the test-time speech query $\boldsymbol{\mathrm{a}}$ nor any images in $\mathcal{M}$ occur in the support set.
This image-speech task was considered in~\cite{eloff_multimodal_2019,nortje_unsupervised_2020,nortje_direct_2021}, but here, for the first time, we use natural images instead of isolated digit images.
In contrast to the task described next, this is conventional few-shot classification where the model only needs to correctly distinguish between classes seen in the support set, i.e.\ there are no other background or imposter classes.

\subsection{Visually grounded few-shot retrieval}
\label{subsec:retrieval_task}
In contrast, in this task the goal is to test whether a model can search through a large collection of images and retrieve those that depict a few-shot speech query, i.e.\ the matching set $\mathcal{M}$ in this case contains images that depict the $L$ few-shot classes but also images that depict other classes.
These additional images might contain completely unseen classes, or background classes potentially seen during pretraining of the few-shot model.
The model is penalised if it retrieves one of these imposter images.
This few-shot retrieval task was proposed in~\cite{miller_tyler_exploring_2022}.
Their interest was specifically in measuring catastrophic forgetting. 
Since their task requires a model to distinguish between few-shot classes and other classes, it can be used to not only determine whether models can be updated to learn new classes from only a few examples, but also how well the model remembers previously learned (background) classes.
We do not explicitly focus on the catastrophic forgetting problem, but we want to compare to~\cite{miller_tyler_exploring_2022}. %
Therefore, we also consider this retrieval task.

For both tasks we need a distance metric $D_\mathcal{S}(\boldsymbol{\mathrm{a}},\boldsymbol{\mathrm{v}})$ between instances from the speech and vision modalities. 
Next we describe the model that we use to compute this distance metric.

\section{Multimodal few-shot attention}
\label{sec:model}

Our approach for determining $D_\mathcal{S}(\boldsymbol{\mathrm{a}},\boldsymbol{\mathrm{v}})$ relies on two core components: a model with a word-to-image attention mechanism and a method to mine pairs using a few ground truth word-image examples (given in the support set).

\subsection{Word-to-image attention mechanism}
\label{subsec:model}
\begin{figure*}[tb]
	\centering
	\includegraphics[width=0.9\linewidth]{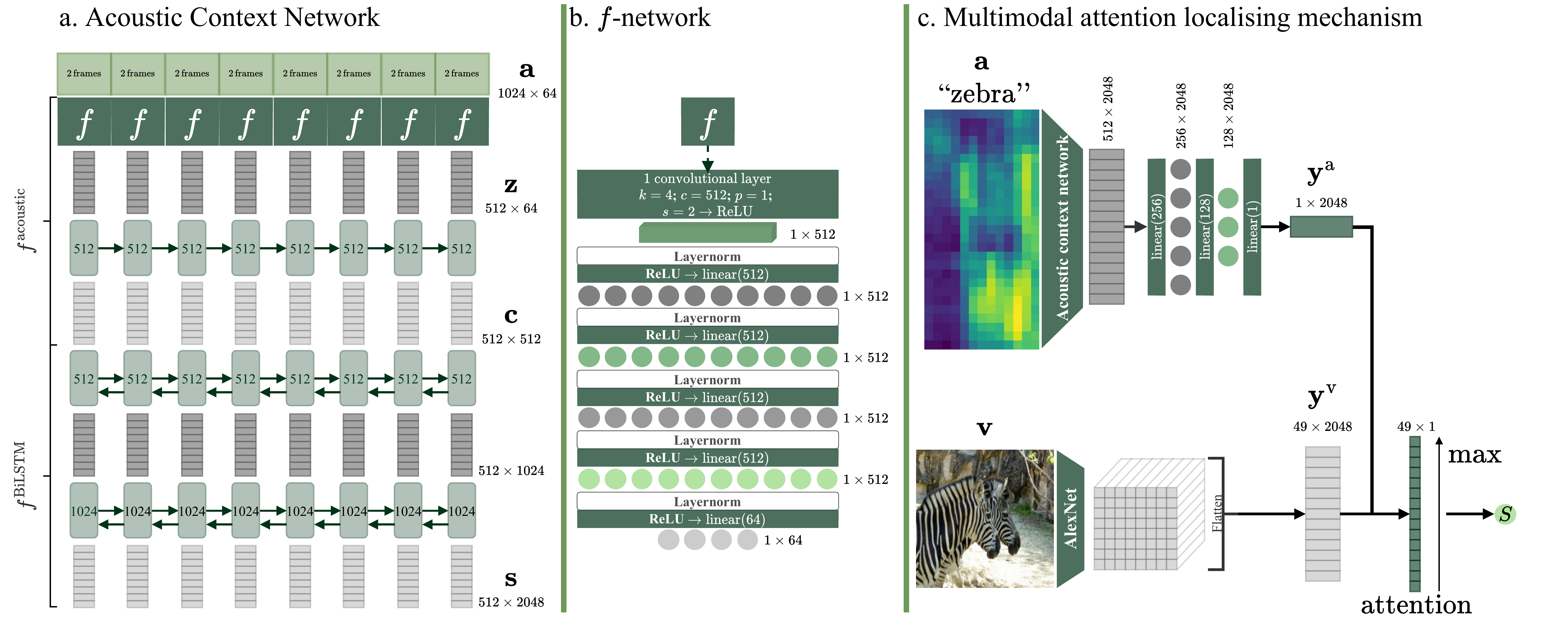}
	\caption{\smallmodel\ consists of (c) a vision and an audio network. The audio network consists of (a + b) an acoustic context network and a BiLSTM network. The audio and vision networks are connected with a word-to-image attention mechanism.}
	\label{fig:model}
\end{figure*}

Our model is shown in Fig.~\ref{fig:model} and we call it \model\ (\textsc{M}ultimodal \textsc{att}ention \textsc{Net}work).
We adapt the multimodal localising attention model of \cite{nortje_towards_2022} that consists of an audio and a vision branch.
For the vision branch, we replace ResNet50~\cite{he_deep_2016} with an adaption of AlexNet~\cite{krizhevsky_imagenet_2017} to encode an image input $\boldsymbol{\mathrm{v}}$ into a sequence of embeddings $\boldsymbol{\mathrm{y}}^{\textrm{v}}$.
Originally~\cite{nortje_visually_2023}, additional linear layers were used after the image embeddings, but removing these did not impact performance.
For the audio branch, we use the same audio subnetwork as \cite{nortje_towards_2022} that consists of an acoustic network $f^{\textrm{acoustic}}$ which extracts speech features from a spoken input $\boldsymbol{\mathrm{a}}$.
However, \cite{nortje_towards_2022} takes an entire spoken utterance as $\boldsymbol{\mathrm{a}}$, whereas we use a single isolated spoken word.
We also add a few linear layers to the BiLSTM network $f^{\textrm{BiLSTM}}$ to encode the speech features into a single audio embedding $\boldsymbol{\mathrm{y}}^{\textrm{a}}$, similar to acoustic word embeddings~\cite{kamper_truly_2019, chung_unsupervised_2016, wang_segmental_2018, holzenberger_learning_2018}.
We connect the vision and audio branches with a multimodal attention mechanism to compare the word embedding $\boldsymbol{\mathrm{y}}^{\textrm{a}}$ to each embedding in $\boldsymbol{\mathrm{y}}^{\textrm{v}}$.

To get this word-to-image attention mechanism, we take the keyword localising attention mechanism of \cite{nortje_towards_2022} which detects whether certain keywords occur in both spoken utterances and images.
However, we aim to only detect whether a single isolated spoken word occurs somewhere within an image.
More specifically, we calculate attention weights over the image embeddings by calculating the dot product between $\boldsymbol{\mathrm{y}}^{\textrm{a}}$ and each embedding in $\boldsymbol{\mathrm{y}}^{\textrm{v}}$.
By taking the maximum over the attention scores, we get a similarity score $S$. %
The higher $S$, the more probable it is that the spoken word corresponds to one or more objects in the image.
If $S$ is low, it is less probable that any object in the image corresponds to the spoken word.

We train \model\ with a contrastive loss: %
\begin{equation}\scriptsize\begin{split}		
		l &= \textrm{MSE}\left(S(\boldsymbol{\mathrm{a}}, \boldsymbol{\mathrm{v}}), 100\right) + \sum_{i=1}^{N_\textrm{pos}}\textrm{MSE}\Bigg(\Big[S(\boldsymbol{\mathrm{a}}, \boldsymbol{\mathrm{v}}_i^+), S(\boldsymbol{\mathrm{a}}_i^+, \boldsymbol{\mathrm{v}})\Big], 100\Bigg)\\
		&+\sum_{i=1}^{N_\textrm{neg}}\textrm{MSE}\Bigg(\Big[S\Big(\boldsymbol{\mathrm{a}}^{-}_{i}, \boldsymbol{\mathrm{v}}), S(\boldsymbol{\mathrm{a}}, \boldsymbol{\mathrm{v}}_i^-), S(\boldsymbol{\mathrm{a}}, \boldsymbol{\mathrm{v}}^{\scriptscriptstyle\textrm{bg}}_{i})\Big], 0\Bigg),
		\label{eq:loss}
\end{split}\end{equation}
where $S$ is calculated with \model\ and we limit $S\in[0, 100]$.
Intuitively this loss should push $\boldsymbol{\mathrm{a}}$, $\boldsymbol{\mathrm{v}}$ and the positive examples $\boldsymbol{\mathrm{a}}^+_i$ and $\boldsymbol{\mathrm{v}}^+_i$ closer
together using a mean square error (MSE) that pushes the list of similarities to $100$.
At the same time the loss %
should push the negative examples $\boldsymbol{\mathrm{a}}^{-}_{i}$, $\boldsymbol{\mathrm{v}}^{-}_{i}$ and $\boldsymbol{\mathrm{v}}^{\scriptscriptstyle\textrm{bg}}_{i}$ away from these positives (through the MSE to 0).
But before we can do this, we need positive ($\boldsymbol{\mathrm{a}}_{i}^+$, $\boldsymbol{\mathrm{v}}_{i}^+$), 
negative ($\boldsymbol{\mathrm{a}}^{-}_{i}$, $\boldsymbol{\mathrm{v}}^{-}_{i}$, $\boldsymbol{\mathrm{v}}^{\scriptscriptstyle\textrm{bg}}_{i}$) and anchor ($\boldsymbol{\mathrm{a}}$, $\boldsymbol{\mathrm{v}}$) pairs. 

\subsection{Few-shot pair mining}
\label{subsec:mining}
For few-shot training, we only have the small number of ground truth examples %
in the support set $\mathcal{S}$.
This would not be sufficient to train the model. %
To overcome this, \cite{nortje_direct_2021} proposed a pair mining scheme:
use the audio examples in $\mathcal{S}$ and compare each example to each utterance in a large collection of unlabelled audio utterances, and similarly for the images.
The mined items can then be used to construct more word-image pairs for training.
While in~\cite{nortje_direct_2021} the unlabelled collection of audio consisted of isolated spoken words (which was artificially segmented), here we consider an unlabelled collection of audio consisting of full spoken utterances (a more realistic scenario).

The simple isolated-word comparison approach used in~\cite{nortje_direct_2021} is not adequate for this setting. 
We employ another approach.
We have a %
spoken word in our support set that we want to match to unlabelled unsegmented utterances in a large audio collection. 
This is similar to fuzzy string search, i.e.\ finding a set of strings that approximately match a given pattern.
However, algorithms from string search are not directly applicable to speech since they operate on a discrete alphabet. 
We therefore use \qbert\ (query-by-example with HuBERT). 
The idea is to encode speech as a set of discrete units that approximate phones. 
Then we can apply standard string search algorithms to find examples that match a given query word.
We use HuBERT \cite{hsu_hubert_2021} to map input speech into discrete units. 
Concretely, we use layer seven of HuBERT-Base for K-means quantisation with 100 clusters.
Then we divide the units into variable-duration phone-like segments following \cite{kamper_towards_2021}.
Finally, we search the dataset by aligning the query to each utterance using the Needleman-Wunsch algorithm \cite{needleman_general_1970}.
An alternative to \qbert\ would have been to use %
dynamic time warping (DTW), %
as is done in~\cite{nortje_direct_2021}. However, in a developmental experiment we found that DTW achieves an isolated word retrieval $F_1$ score of  76.8\% while \qbert\ achieves 98.7\%.

Using \qbert, we compare each spoken utterance in an unlabelled collection of audio utterances to each spoken word example in $\mathcal{S}$.
For each utterance, we take the highest score across the $K$ word examples per class and rank the utterances from highest to lowest for each class. 
The first $n$ utterances with the highest scores for a class are predicted to contain the spoken form of the word.
Additionally, we use \qbert's predicted word segments to isolate matched words. %
To mine image pairs, we follow the same steps, but instead we use AlexNet~\cite{krizhevsky_imagenet_2017} to extract a single embedding for each image and use cosine distance to compare image embeddings to one another.
To get word-image pairs, we mine an image from the same predicted class as a segmented word.
Negative pairs are taken from the positive pairs of other classes.
We also mine an extra negative image $\boldsymbol{\mathrm{v}}^{\scriptscriptstyle\textrm{bg}}_{i}$ from a set known to not contain any of the few-shot classes (referred to as the background data, see below). 
Therefore, during the few-shot retrieval task, images containing few-shot classes can be distinguished from images that depicts none of the few-shot classes.

\section{English Few-Shot Speech-Image Experiments}
\label{sec:english}

Before we get to an actual low-resource setting, we do experiments and perform analyses on an English benchmark.

\subsection{Experimental setup}
\label{sec:english_setup}

\begin{figure*}[tb]
	\centering
	\includegraphics[width=0.8\linewidth]{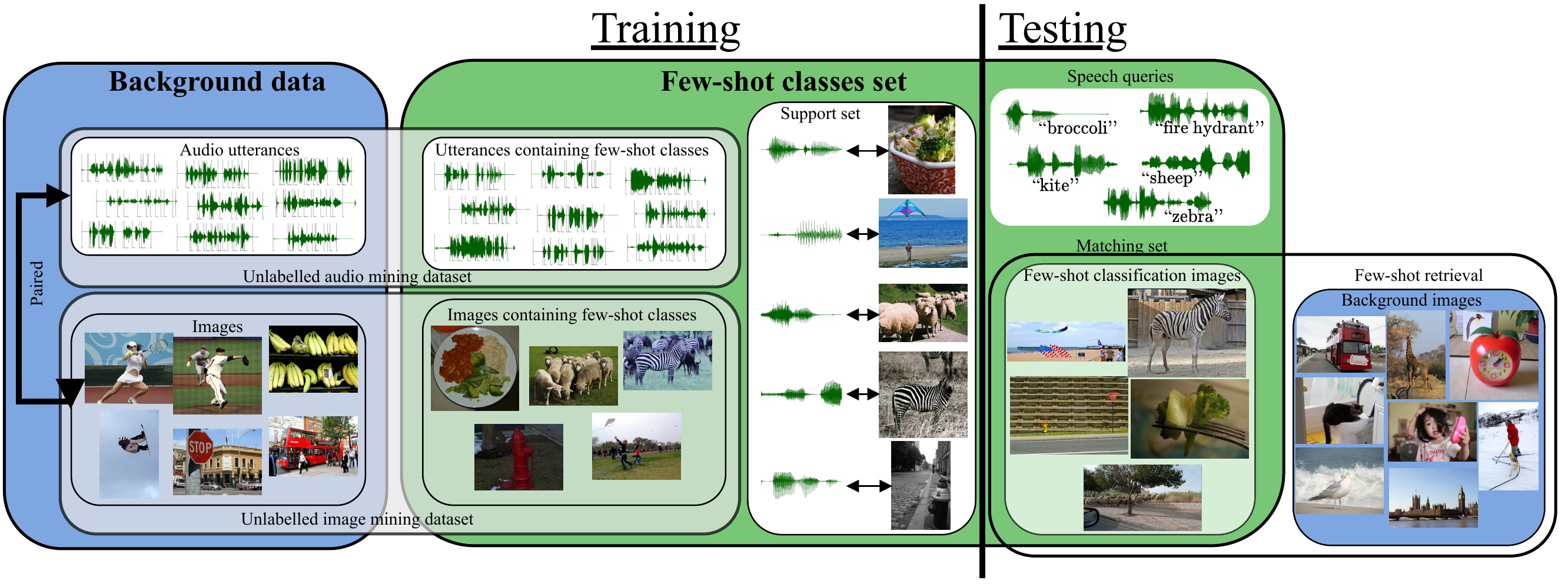}
	\caption{%
         The SpokenCOCO data splits used to train and evaluate the \smallmodel\ model.
         The background data (blue background) consists of spoken audio utterances and images belonging to concepts not present in the support set.
         The mining splits consist of single-modality audio and image samples from which the training data is artificially extended, and include both background samples and samples belonging to the few-shot classes (green).
    }
	\label{fig:spokencoco_fs_data}
\end{figure*}

\subsubsection{Data} 
\label{sec:subsubsec_data}
For our English experiments, we use the SpokenCOCO Corpus~\cite{hsu_text-free_2021} which  consists of the MSCOCO~\cite{lin_microsoft_2014} images with recorded spoken captions corresponding to the MSCOCO textual captions.
Fig.~\ref{fig:spokencoco_fs_data} shows how we partition the dataset.
Firstly, we use the setup of \cite{miller_tyler_exploring_2022} to divide the SpokenCOCO dataset into a few-shot set (green) and a background set not containing any of the few-shot classes  (blue).\footnote{\cite{miller_tyler_exploring_2022} refers to classes occurring in the background data as base classes.}
We use the same few-shot classes as \cite{miller_tyler_exploring_2022}: \textit{brocolli}, \textit{fire hydrant}, \textit{kite}, \textit{sheep} and \textit{zebra}.
The background data is used to pretrain \model (\S\ref{sec:models}).

For the few-shot set, we further divide it into training and testing according to the splits used by \cite{miller_tyler_exploring_2022}.
We use this testing set to sample isolated spoken word queries and matching images for testing. 
For the few-shot classification task (\S\ref{subsec:classification_task}), we sample images only from the few-shot test set.
Since the few-shot retrieval task (\S\ref{subsec:retrieval_task}) requires a single large image matching set, we take all of the images in the few-shot test set as well as the images in the background test set.
We sample the support set $\mathcal{S}$ from the few-shot training set (\S\ref{sec:task}), using the Montreal forced aligner~\cite{mcauliffe_montreal_2017} to isolate the few-shot words.
To mine pairs (\S\ref{subsec:mining}), we need both an unlabelled audio and image dataset to mine pairs from; for this we use the remainder of the few-shot training data that does not include the support set as well as the background training data, shown in the left half of Fig.~\ref{fig:spokencoco_fs_data}.
From these unlabelled collections, we mine pairs: the $n=600$ highest ranking examples per class (\S\ref{subsec:mining}).
Lastly, these pairs are split into training and validation pairs.

Utterances are parametrised as mel-spectograms with a hop length of 10~ms, a window of 25~ms and 40~mel bins. 
These are truncated or zero-padded to 1024 frames. 
Images are resized to 224$\times$224 pixels and normalised with means and variances calculated on ImageNet~\cite{deng_imagenet_2009}.

\subsubsection{Models}
\label{sec:models}

Fig.~\ref{fig:model} illustrates our model, \model\ (\S\ref{subsec:model}).
For the image branch, we use an adaption of AlexNet~\cite{krizhevsky_imagenet_2017} to get image embeddings. 
This image branch is also initialised using the pretrained convolutional encoder (before the classification network) of AlexNet.
We use an adaption of \cite{nortje_towards_2022}'s audio network for the audio branch.
This acoustic network is pretrained on LibriSpeech~\cite{panayotov_librispeech_2015} and the multilingual (English and Hindi) Places dataset~\cite{harwath_vision_2018}
using a self-supervised contrastive predictive coding task~\cite{van_niekerk_vector-quantized_2020}.
While there are more modern alternatives for the vision and audio networks, we chose these particular variants to limit computational requirements.
After initialisation, %
the combined \model\ model is pretrained on the background data (blue, Fig.~\ref{fig:spokencoco_fs_data}) using the contrastive speech-image retrieval loss of \cite{harwath_unsupervised_2016}.
This is then the starting point for the model that we update using mining.

\begin{table}[tb]
    \newcommand{\cmark}{\color{black}\ding{51}}%
    \newcommand{\xmark}{\color{gray!50}\ding{55}}%
    \caption{The setup for each \smallmodel\ variant.}
    \captionsep
	\label{tab:model_names}
	\centering
	\footnotesize
	\setlength{\tabcolsep}{0.35em}
	\renewcommand{\arraystretch}{1}
	\begin{tabular*}{\linewidth}{ l@{\extracolsep{\fill}}ccc }
		\toprule
		Model & Pretrain & Mine training pairs & Fine-tune\\
		\midrule  
		\smallmodel & \cmark & \cmark & \cmark\\ %
		\smallmodel, no mining & \cmark & \xmark & \cmark\\ %
		\smallmodel, no fine-tuned & \cmark & \xmark & \xmark\\ %
		\bottomrule
	\end{tabular*}
\end{table}

During training on the SpokenCOCO mined pairs, we take $N_\textrm{pos}=5$ and $N_\textrm{neg}=11$ in~\eqref{eq:loss}.
These values were fine-tuned on the validation pairs.
We train all models with Adam~\cite{kingma_adam_2015} for 100 epochs using a validation task for early stopping. %
For the validation task, we use the validation set to get one positive image $\boldsymbol{\mathrm{v}}^+$ and one negative image $\boldsymbol{\mathrm{v}}^-$ for each validation word-image $(\boldsymbol{\mathrm{a}},\, \boldsymbol{\mathrm{v}})$ pair. 
The validation task measures whether the model will place $\boldsymbol{\mathrm{v}}$ and $\boldsymbol{\mathrm{v}}^+$ closer to  $\boldsymbol{\mathrm{a}}$ than it would  $\boldsymbol{\mathrm{v}}^-$. 

To understand the models we discuss in the following section, we give the model names and their setup in Table~\ref{tab:model_names}.
Our full model is referred to as ``\model''; it is pretrained on background data and then fine-tuned on the mined pairs.
For the ``\model, no mining'' model, %
instead of training on the mined pairs, we only update the model on the samples in the support set.
The ``\model, no fine-tuned'' model consists of only the pretrained model; it is not updated with the few-shot classes in any way.
To do the few-shot tasks with this model, we use the indirect few-shot method of \cite{eloff_multimodal_2019, nortje_unsupervised_2020}:
each $\boldsymbol{\mathrm{a}}$ is compared to each $\boldsymbol{\mathrm{a}}_{j}$ in $\mathcal{S}$ to find the audio example closest to the query.
The image $\boldsymbol{\mathrm{v}}_{j}$ corresponding to the closest $\boldsymbol{\mathrm{a}}_{j}$ is then used to calculate the similarity to each image $\boldsymbol{\mathrm{v}}_{i}$ in $\mathcal{M}$.
This model can thus be seen as evaluating the quality of the embedding spaces obtained through pertaining when data from unseen classes are presented to the model.

\subsubsection{Few-shot evaluation tasks} 
\label{subsubsec:few-shot_tasks_implementation}
We evaluate our approach on two tasks (as explained in \S\ref{sec:task}): a traditional few-shot classification task and a few-shot retrieval task.
For both tasks, the $K$-shot $L$-way support set $\mathcal{S}$ contains $K$ ground truth spoken word-image pairs for each of the $L=5$ classes and is used to mine pairs for training and validation.
In the few-shot classification task, we sample $1000$ episodes where each episode contains $L$ spoken word queries $\boldsymbol{\mathrm{a}}$, one for each class, and a matching set $\mathcal{M}$ which contains one image $\boldsymbol{\mathrm{v}}_{i}$ for each class.   
In the few-shot retrieval task, instead of having one image per class, $\mathcal{M}$ consists of $5000$ images $\boldsymbol{\mathrm{v}}_{i}$ where some depict a few-shot class and others do not. 
Here, 20 query words are taken per class and averaged to get $\boldsymbol{\mathrm{a}}$. 
For each of the $L$ queries $\boldsymbol{\mathrm{a}}$, these $5000$ images are ranked from highest to lowest similarity.
The precision at $N$ ($P@N$) score is the proportion of images in the top $N$ highest ranking images that are from the same class as $\boldsymbol{\mathrm{a}}$.
$N$ is the actual number of images in $\mathcal{M}$ that depicts the word class. 

\subsection{Experimental results}
\label{sec:english_results}

\begin{table}[!b]
	\caption{%
		$P$@N few-shot retrieval scores (\%) on the five few-shot classes. $K$ is the number of support-set examples per class.	}
  \captionsep
	\label{tab:retrieval_task}
	\centering
	\footnotesize
	\setlength{\tabcolsep}{0.25em}
	\renewcommand{\arraystretch}{1.2}
	\begin{tabular*}{\linewidth}{ l@{\extracolsep{\fill}}cccc }
		\toprule
		\multicolumn{1}{c}{Model} & 
		\multicolumn{4}{c}{$K$}\\
		\cline{2-5}
		 & 5 & 10 & 50 & 100 \\
		\midrule
		Naive fine-tuned~\cite{miller_tyler_exploring_2022} & -- & -- & -- & \textbf{52.5}\\
		Oracle masking~\cite{miller_tyler_exploring_2022} & -- & 8.4$\pm$0.0  & 24.0$\pm$0.1 & 35.5$\pm$0.2\\
    \addlinespace		
    \smallmodel\ & \textbf{40.3$\pm$0.1} & \textbf{44.2$\pm$0.1} & \textbf{41.7$\pm$0.2} & 43.7$\pm$0.1\\
    \smallmodel, no mining & 13.2$\pm$0.6 & 34.8$\pm$0.7 & 40.9$\pm$0.3 & 40.5$\pm$0.5 \\
		\smallmodel, no fine-tuned\hspace{0.5em} & 22.0$\pm$0.4 & 24.1$\pm$0.8 & 22.7$\pm$0.5 & 23.2$\pm$1.1\\
		\bottomrule
	\end{tabular*}
\end{table}

\begin{table}[!b]
	\caption{%
		Few-shot word classification accuracy  (\%) when varying
  the number of shots per class $K$.}
  \captionsep
	\label{tab:few-shot_classification}
	\centering
	\footnotesize
	\setlength{\tabcolsep}{1em}
	\renewcommand{\arraystretch}{1.2}
	\begin{tabular*}{\linewidth}{ l@{\extracolsep{\fill}}cccc }
		\toprule
		\multicolumn{1}{c}{Model} & 
		\multicolumn{4}{c}{$K$}\\
		\cline{2-5}
		 & 5 & 10 & 50 & 100 \\
		\midrule
            \smallmodel\ & \textbf{80.1} & \textbf{81.1} & 88.5 & 93.2\\
\smallmodel, no mining %
        & 53.1 & 79.4 & \textbf{93.3} & \textbf{95.5}\\  
		\smallmodel, no fine-tuned & 50.4 & 48.0 & 48.5 & 47.7\\
		\bottomrule
	\end{tabular*}
\end{table}

We start by comparing to
two of \cite{miller_tyler_exploring_2022}'s models on the few-shot retrieval task.
The first %
is their naive model, which is pretrained on background classes and then fine-tuned on $K=100$ examples for each of the $L=5$ classes.
The second is an oracle masking model in which the contrastive loss used during fine-tuning ensures that a negative image does not contain any instance of the anchor few-shot class.
The results are given in Table~\ref{tab:retrieval_task}.%
\footnote{These scores are slightly different from those in \cite{nortje_visually_2023} because we do not apply linear layers over the image encodings (\S\ref{subsec:model}).
} 
(Not all settings considered here were evaluated in~\cite{miller_tyler_exploring_2022}, so these are indicated with dashes.)

Our full \model\ model %
outperforms the oracle model %
across all values of $K$. 
Neither \model\ nor the oracle masking works as well as the %
naive fine-tuned approach for a high number of shots (line 1, $K = 100$).
In the forth line we see that our no mining approach, which is equivalent to naive fine-tuning,
does worse than the naive model from \cite{miller_tyler_exploring_2022}
at $K = 100$.
It is important to note that we use a different architecture.
We can, however, conclude that direct fine-tuning %
only works with a large number of shots; as the number of shots increases, we get closer to a standard supervised learning setting, and it is therefore unsurprising that at some point naive fine-tuning starts to outperform few-shot methods.
But, taking all this together, it is clear that our approach outperforms the existing methods with fewer shots.

To determine the contribution of both mining and fine-tuning, we do an experiment where we do not update \model\ on the 
few-shot classes after pretraining it on the background data (``\model, no fine-tuned''). 
We see that these two components improve the scores by roughly 20\% in absolute performance when comparing lines~3 and 5 in Table~\ref{tab:retrieval_task}.

To further analyse the performance gains from mining, we now consider the conventional few-shot word classification task (\S\ref{subsec:classification_task}). 
This task wasn't used in~\cite{miller_tyler_exploring_2022}.
Table~\ref{tab:few-shot_classification} shows that the few-shot classification scores increase as $K$ increases when we use mined pairs.
For the no fine-tuning method, the scores are lower than that of \model\ and decrease slightly as $K$ increases.
The no mining approach in which we only update the model on the support set samples has a steep %
increase in scores from $K=5$ to $K=10$, after which the scores overtake \model.
Again, this makes sense as the amount of training data ($K$) increases.
Again, these results illustrate the effectiveness of \model\ when we have fewer shots.

Altogether, we set a competitive multimodal baseline for both few-shot retrieval and word classification in settings where the number of shots is small.
However, the results do raise some questions, including the following: why do the retrieval scores for \model\ plateau (Table~\ref{tab:retrieval_task}) but the classification scores increase (Table~\ref{tab:few-shot_classification}) as $K$ increases?
We unpack this as part of the analysis in the next section.

\section{Further Analysis of English}
\label{sec:english_ablation}

In this section we investigate the retrieval performance plateau we encountered in the previous section. 
We also present a finer-grained analysis looking into various aspects that contribute to our approach's performance.

\subsection{Qualitative error analysis}

\begin{figure*}[tb]
    	\setlength{\tabcolsep}{2pt}
    	\newcommand{\cmark}{\color{indiagreen}\ding{51}}%
    	\newcommand{\xmark}{\color{red}\ding{55}}%
    	\centering
        \def\imgsize{1.5cm}
        \def\imgheight{0.75cm}
        \footnotesize
    	\begin{tabular}{ccccc|ccccc}
    		\multicolumn{5}{c|}{\color{gray} query: \it fire hydrant} &
    		\multicolumn{5}{c }{\color{gray} query: \it kite} \\
    		\multicolumn{5}{c|}{\includegraphics[width=\imgsize, height=\imgheight]{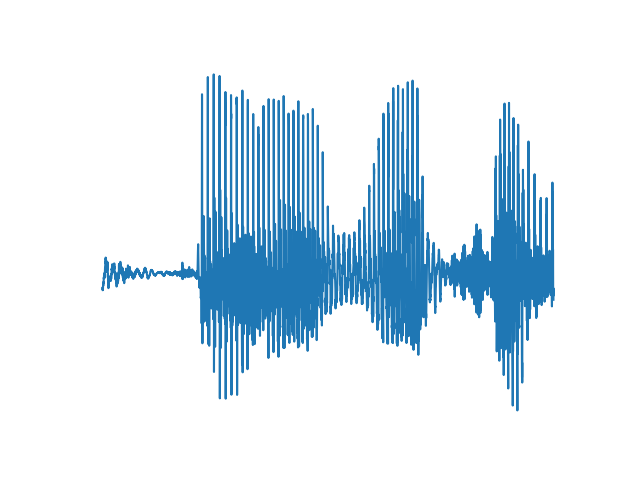}} &
    		\multicolumn{5}{c }{\includegraphics[width=\imgsize, height=\imgheight]{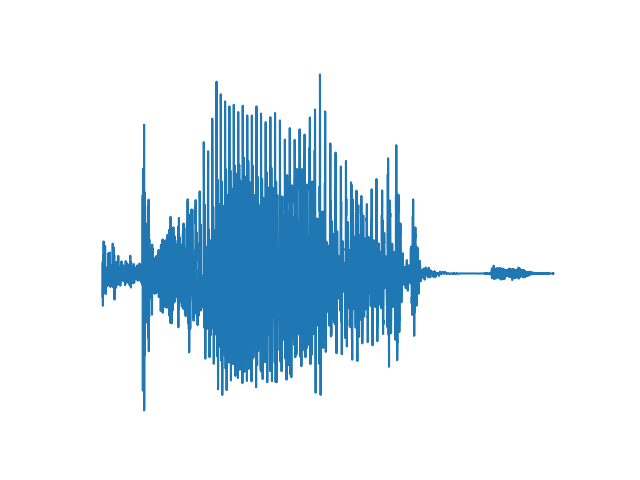}} \\
    		\multicolumn{5}{c|}{\color{gray} task: \it few-shot retrieval} &
    		\multicolumn{5}{c }{\color{gray} task: \it few-shot retrieval} \\
    		rank: 1 · \cmark &
    		rank: 2 · \xmark &
    		rank: 3 · \xmark &
    		rank: 4 · \xmark &
    		rank: 5 · \xmark &
            rank: 1 · \cmark &
    		rank: 2 · \xmark &
    		rank: 3 · \xmark &
    		rank: 4 · \cmark &
    		rank: 5 · \cmark \\
            \includegraphics[width=\imgsize, height=\imgsize]{imgs/100-loc-v2-ret/COCO_val2014_000000553511.jpg} &
            \includegraphics[width=\imgsize, height=\imgsize]{imgs/100-loc-v2-ret/COCO_val2014_000000209222.jpg} &
            \includegraphics[width=\imgsize, height=\imgsize]{imgs/100-loc-v2-ret/COCO_val2014_000000261796.jpg} &
            \includegraphics[width=\imgsize, height=\imgsize]{imgs/100-loc-v2-ret/COCO_val2014_000000361142.jpg} &
            \includegraphics[width=\imgsize, height=\imgsize]{imgs/100-loc-v2-ret/COCO_val2014_000000546823.jpg} &
            \includegraphics[width=\imgsize, height=\imgsize]{imgs/100-loc-v2-ret/COCO_val2014_000000074646.jpg} &
            \includegraphics[width=\imgsize, height=\imgsize]{imgs/100-loc-v2-ret/COCO_val2014_000000055299.jpg} &
            \includegraphics[width=\imgsize, height=\imgsize]{imgs/100-loc-v2-ret/COCO_val2014_000000457559.jpg} &
            \includegraphics[width=\imgsize, height=\imgsize]{imgs/100-loc-v2-ret/COCO_val2014_000000527528.jpg} &
            \includegraphics[width=\imgsize, height=\imgsize]{imgs/100-loc-v2-ret/COCO_val2014_000000576654.jpg} \\
            score: 94.0 &
    		score: 80.7 &
    		score: 77.0 &
    		score: 74.0 &
    		score: 72.2 &
    		score: 103.8 &
    		score:  99.2 &
    		score:  95.2 &
    		score:  95.2 &
    		score:  91.8 \\ [10pt]
    		\multicolumn{5}{c|}{\color{gray} task: \it few-shot classification} &
    		\multicolumn{5}{c }{\color{gray} task: \it few-shot classification} \\
    		broccoli &
    		zebra &
    		\bf fire hydrant &
    		sheep &
    		kite &
            broccoli &
    		zebra &
    		fire hydrant &
    		sheep &
    		\bf kite \\
    		\includegraphics[width=\imgsize, height=\imgsize]{imgs/100-clf-fire-hydrant-0/COCO_val2014_000000559707.jpg} &
    		\includegraphics[width=\imgsize, height=\imgsize]{imgs/100-clf-fire-hydrant-0/COCO_val2014_000000426297.jpg} &
    		\includegraphics[width=\imgsize, height=\imgsize]{imgs/100-clf-fire-hydrant-0/COCO_val2014_000000580757.jpg} &
    		\includegraphics[width=\imgsize, height=\imgsize]{imgs/100-clf-fire-hydrant-0/COCO_val2014_000000461405.jpg} &
    		\includegraphics[width=\imgsize, height=\imgsize]{imgs/100-clf-fire-hydrant-0/COCO_val2014_000000224675.jpg} &
            \includegraphics[width=\imgsize, height=\imgsize]{imgs/100-clf-fire-hydrant-0/COCO_val2014_000000559707.jpg} &
    		\includegraphics[width=\imgsize, height=\imgsize]{imgs/100-clf-fire-hydrant-0/COCO_val2014_000000426297.jpg} &
    		\includegraphics[width=\imgsize, height=\imgsize]{imgs/100-clf-fire-hydrant-0/COCO_val2014_000000580757.jpg} &
    		\includegraphics[width=\imgsize, height=\imgsize]{imgs/100-clf-fire-hydrant-0/COCO_val2014_000000461405.jpg} &
    		\includegraphics[width=\imgsize, height=\imgsize]{imgs/100-clf-fire-hydrant-0/COCO_val2014_000000224675.jpg} \\
    		score:  1.2 &
    		score: \textminus0.4 &
    		score: 15.5 &
    		score:  2.4 &
    		score:  7.0 &
            score: \textminus6.4 &
    		score: \textminus0.5 &
    		score: 10.2 &
    		score:  7.7 &
    		score: 12.2 \\
    	\end{tabular}
        \caption{%
        Examples of retrieval and few-shot classification for two queries using the $K = 100$ \smallmodel\  model.
        Concepts that associate strongly with context, such as \textit{fire hydrant} which often appears in urban environments, are more challenging to retrieve than to classify.
        }
    	\label{fig:retrieval-vs-classification}
\end{figure*}

We start with a qualitative analysis to compare the few-shot classification and retrieval tasks.
Fig.~\ref{fig:retrieval-vs-classification} shows
the five matching set images for classification and
the five images that the model retrieves when given the spoken queries ``fire hydrant'' and ``kite''. %
The analysis {suggests}
that for some classes the model depends on the contextual information to identify the class. 
Since the few-shot image classes and their contexts are quite distinct from one another, it does not have a significant influence on the classification task (bottom, %
{Fig.~\ref{fig:retrieval-vs-classification}}).
However, this has a more significant effect on retrieval since contextual information may overlap more between the few-shot images and background images (top).
E.g.\ there might be multiple images containing streets, but only some of them contain \textit{fire hydrants}.
For classification, this learned association actually helps since none of the other few-shot classes involves streets.

\subsection{Per-keyword analysis}
\label{subsec:per_keyword_analysis}

\begin{table}[!b]
    \caption{$P@N$ few-shot retrieval scores (\%) for each of the five few-shot classes. $K$ is the number of support-set examples per class. The number of examples present in the matching set $\mathcal{M}$ for each class, $N$, is given in brackets.}
    \captionsep
    \label{tab:keyword_analysis}
    \centering
    \footnotesize
    \renewcommand{\arraystretch}{1.2}
    \begin{tabular*}{\linewidth}{l@{\extracolsep{\fill}}ccccc}
        \toprule
         & \textit{broccoli}  & \textit{fire hydrant}  & \textit{kite}  & \textit{sheep}  & \textit{zebra}\\
         $K$ & (57) & (62) & (91) & (63) & (90)\\
         \midrule
         5 & 40.4 & 11.3 & 27.5 & 33.3 & 77.8\\
         10 & 49.1 & 9.7 & 30.8 & 33.3 & 85.6\\
         50 & 50.9 & 8.1 & 20.9 & 36.5 & 82.2\\
         100 & 52.6 & 4.8 & 33.0 & 38.1 & 80.0\\
         \bottomrule
         \end{tabular*}
    \end{table}

Table~\ref{tab:keyword_analysis} presents the individual retrieval performance for each of the five few-shot keywords. %
(In \S\ref{sec:english_results}, specifically Table~\ref{tab:retrieval_task}, retrieval scores were aggregated over the keyword types.)
We observe a large variance in performance across the concepts:
\textit{fire hydrant} is the most challenging keyword to retrieve (with a performance as low as 4.8\% in $P@N$), while \textit{zebra} is the easiest (over 85\% when $K = 10$).
The differences between keywords could be due to a number of factors,
object size and word frequency probably being among the most important ones, as suggested by \cite[Table~4]{miller_tyler_exploring_2022}. %

Interestingly, if we look at how the performance varies with the number of shots $K$, we get a more nuanced picture than the one provided by the aggregated results (in Table~\ref{tab:retrieval_task}).
While we previously observed the aggregated performance staying approximately the same with larger $K$, here we instead see that there are variations in both directions depending on the keyword: the retrieval scores improve with $K$ for \textit{sheep} and \textit{broccoli}, but degrade for \textit{fire hydrant} and fluctuates for \textit{kite} and \textit{zebra}.
We speculate that as $K$ increases for classes that have low retrieval scores (\textit{fire hydrant}, \textit{kite}), the model becomes more dependent on the recurring contextual (background) features, which are especially difficult to disentangle for the objects that occupy only a small part of the image.
I.e.\ when the few-shot object is small, the model struggles to focus on the few-shot object and therefore depends on the contextual data to learn the class.
Therefore, increasing $K$ does not help.
We have already seen this to some degree in the examples of Fig.~\ref{fig:retrieval-vs-classification}; next we investigate this by analysing the model's localisations.

\subsection{Localisation visualisation}
\label{subsec:localisation}

\begin{figure*}
	\setlength{\tabcolsep}{2pt}
    \def\imgsize{1.5cm}
    \def\imgheight{0.75cm}
    \footnotesize
	\newcommand{\cmark}{\color{indiagreen}\ding{51}}%
	\newcommand{\xmark}{\color{red}\ding{55}}%
	\centering
	\begin{tabular}{cc|cc|cc|cc|cc}
		\multicolumn{2}{c}{\it \color{gray} broccoli} &
		\multicolumn{2}{c}{\it \color{gray} zebra} &
		\multicolumn{2}{c}{\it \color{gray} fire hydrant} &
		\multicolumn{2}{c}{\it \color{gray} sheep} &
		\multicolumn{2}{c}{\it \color{gray} kite} \\
		\multicolumn{2}{c}{\includegraphics[width=\imgsize, height=\imgheight]{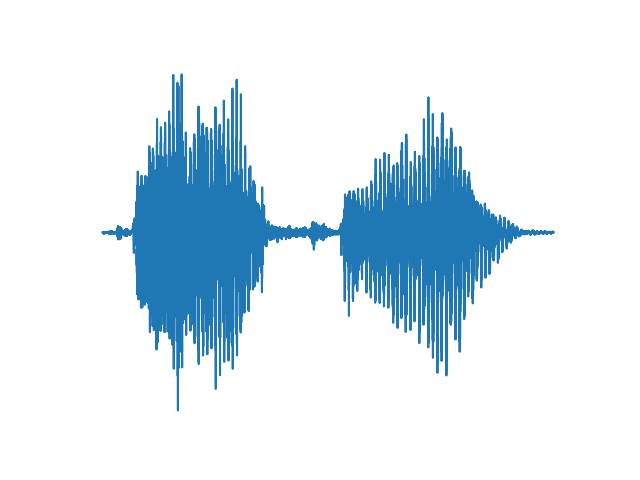}} &
		\multicolumn{2}{c}{\includegraphics[width=\imgsize, height=\imgheight]{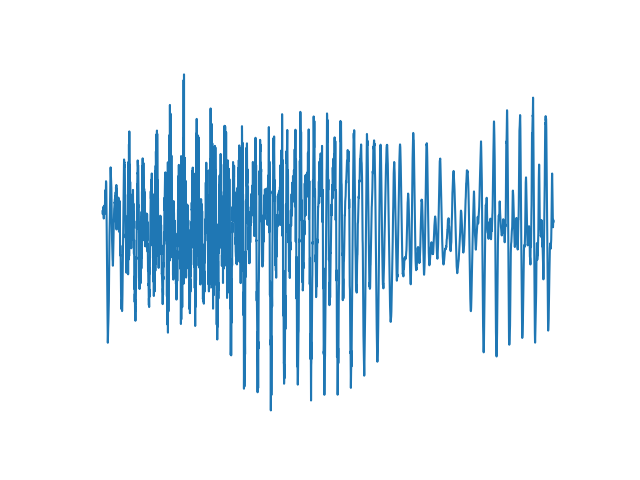}} &
		\multicolumn{2}{c}{\includegraphics[width=\imgsize, height=\imgheight]{imgs/audio-fire-hydrant.png}} &
		\multicolumn{2}{c}{\includegraphics[width=\imgsize, height=\imgheight]{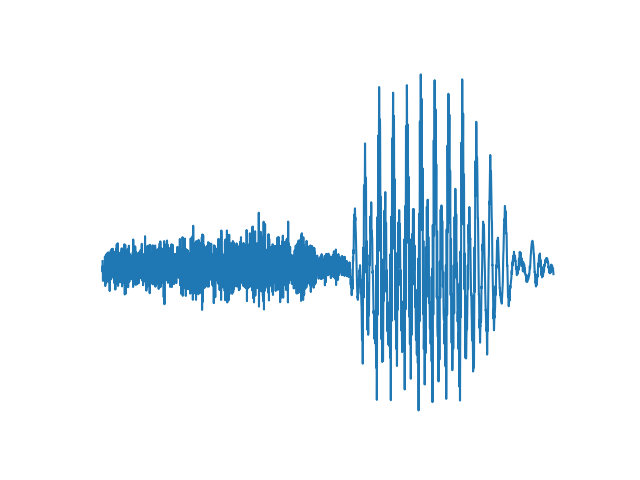}} &
		\multicolumn{2}{c}{\includegraphics[width=\imgsize, height=\imgheight]{imgs/audio-kite.png}} \\
        \multicolumn{2}{c}{\color{gray} IOU: 19.3\%} &
        \multicolumn{2}{c}{\color{gray} IOU: 35.3\%} &
        \multicolumn{2}{c}{\color{gray} IOU:  0.2\%} &
        \multicolumn{2}{c}{\color{gray} IOU: 12.6\%} &
        \multicolumn{2}{c}{\color{gray} IOU: 0.5\%} \\[0.1cm]
        \multicolumn{2}{c|}{rank: 1 · is correct: \xmark} &
        \multicolumn{2}{c|}{rank: 1 · is correct: \cmark} &
        \multicolumn{2}{c|}{rank: 1 · is correct: \cmark} &
        \multicolumn{2}{c|}{rank: 1 · is correct: \xmark} &
        \multicolumn{2}{c }{rank: 1 · is correct: \cmark} \\
        \includegraphics[width=\imgsize, height=\imgsize]{imgs/100-loc-v2-ret/COCO_val2014_000000493799.jpg} &
        \includegraphics[width=\imgsize, height=\imgsize]{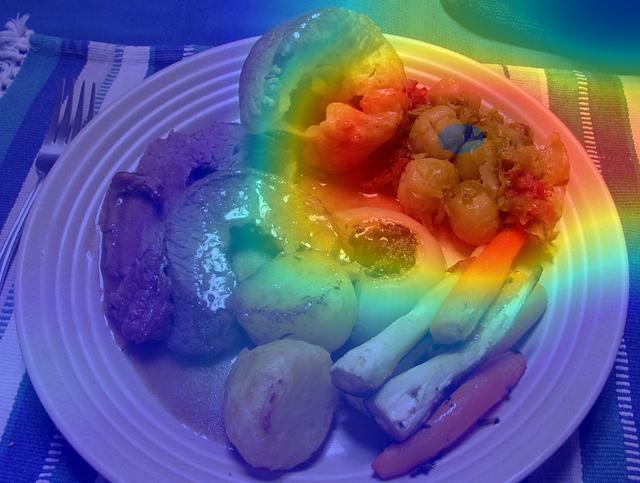} &
        \includegraphics[width=\imgsize, height=\imgsize]{imgs/100-loc-v2-ret/COCO_val2014_000000250758.jpg} &
        \includegraphics[width=\imgsize, height=\imgsize]{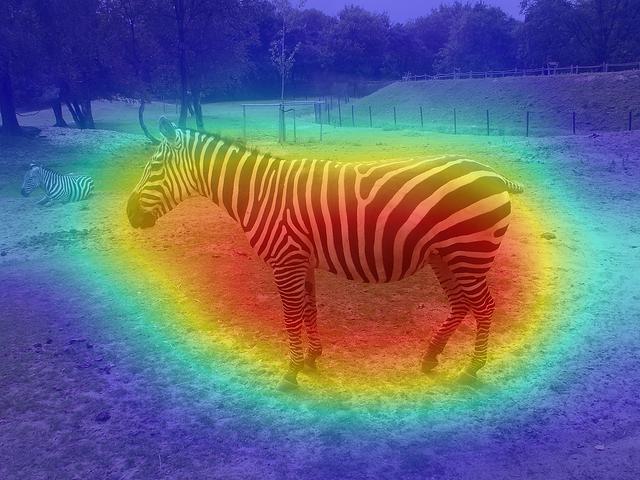} &
        \includegraphics[width=\imgsize, height=\imgsize]{imgs/100-loc-v2-ret/COCO_val2014_000000553511.jpg} &
        \includegraphics[width=\imgsize, height=\imgsize]{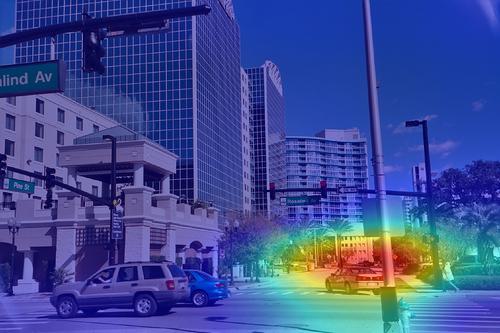} &
        \includegraphics[width=\imgsize, height=\imgsize]{imgs/100-loc-v2-ret/COCO_val2014_000000020247.jpg} &
        \includegraphics[width=\imgsize, height=\imgsize]{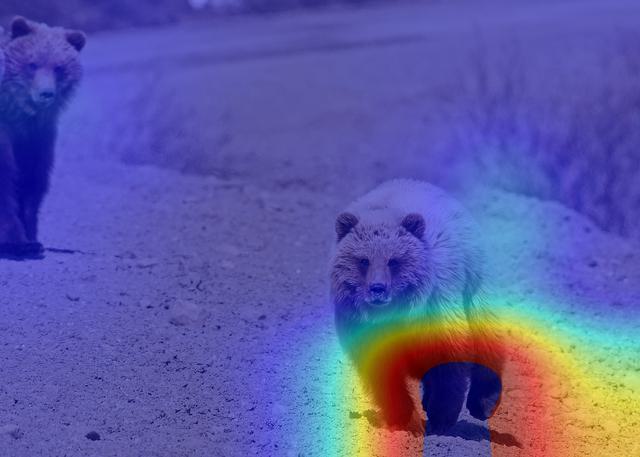} &
        \includegraphics[width=\imgsize, height=\imgsize]{imgs/100-loc-v2-ret/COCO_val2014_000000074646.jpg} &
        \includegraphics[width=\imgsize, height=\imgsize]{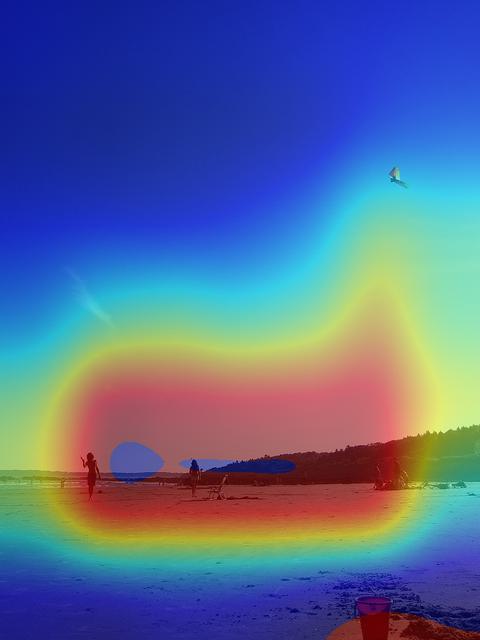} \\
        \multicolumn{2}{c|}{rank: 2 · is correct: \cmark} &
        \multicolumn{2}{c|}{rank: 2 · is correct: \cmark} &
        \multicolumn{2}{c|}{rank: 2 · is correct: \xmark} &
        \multicolumn{2}{c|}{rank: 2 · is correct: \xmark} &
        \multicolumn{2}{c }{rank: 2 · is correct: \xmark} \\
        \includegraphics[width=\imgsize, height=\imgsize]{imgs/100-loc-v2-ret/COCO_val2014_000000092660.jpg} &
        \includegraphics[width=\imgsize, height=\imgsize]{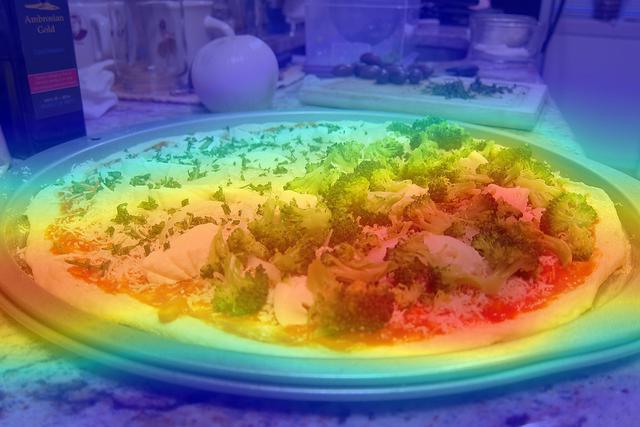} &
        \includegraphics[width=\imgsize, height=\imgsize]{imgs/100-loc-v2-ret/COCO_val2014_000000546011.jpg} &
        \includegraphics[width=\imgsize, height=\imgsize]{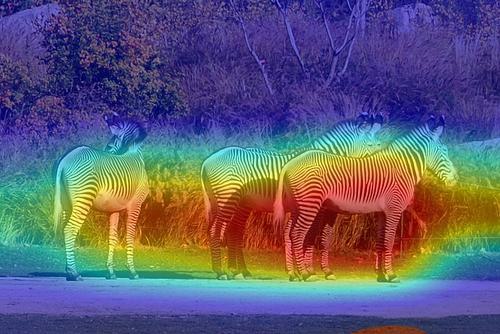} &
        \includegraphics[width=\imgsize, height=\imgsize]{imgs/100-loc-v2-ret/COCO_val2014_000000209222.jpg} &
        \includegraphics[width=\imgsize, height=\imgsize]{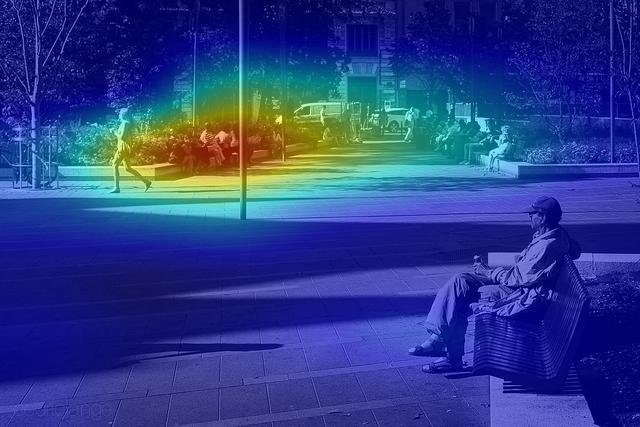} &
        \includegraphics[width=\imgsize, height=\imgsize]{imgs/100-loc-v2-ret/COCO_val2014_000000547519.jpg} &
        \includegraphics[width=\imgsize, height=\imgsize]{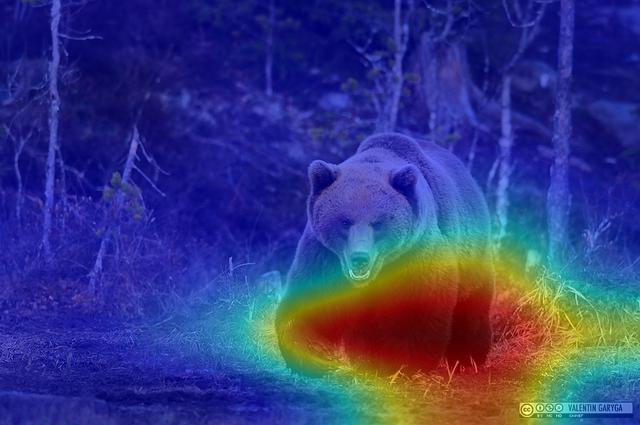} &
        \includegraphics[width=\imgsize, height=\imgsize]{imgs/100-loc-v2-ret/COCO_val2014_000000055299.jpg} &
        \includegraphics[width=\imgsize, height=\imgsize]{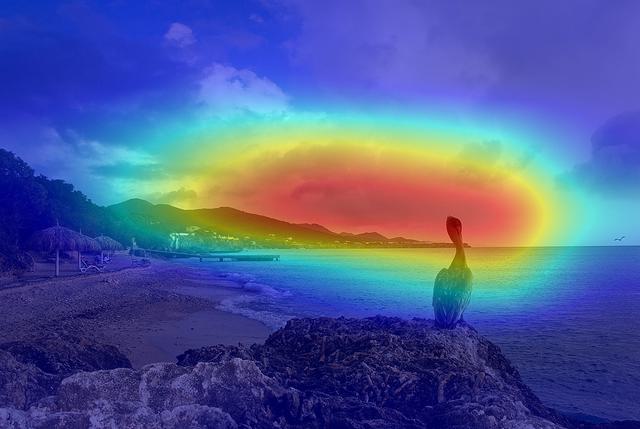} \\
        \multicolumn{2}{c|}{rank: 3 · is correct: \xmark} &
        \multicolumn{2}{c|}{rank: 3 · is correct: \cmark} &
        \multicolumn{2}{c|}{rank: 3 · is correct: \xmark} &
        \multicolumn{2}{c|}{rank: 3 · is correct: \cmark} &
        \multicolumn{2}{c }{rank: 3 · is correct: \xmark} \\
        \includegraphics[width=\imgsize, height=\imgsize]{imgs/100-loc-v2-ret/COCO_val2014_000000100582.jpg} &
        \includegraphics[width=\imgsize, height=\imgsize]{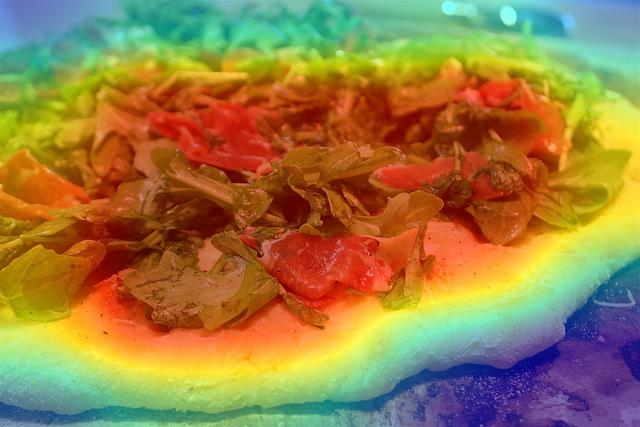} &
        \includegraphics[width=\imgsize, height=\imgsize]{imgs/100-loc-v2-ret/COCO_val2014_000000215245.jpg} &
        \includegraphics[width=\imgsize, height=\imgsize]{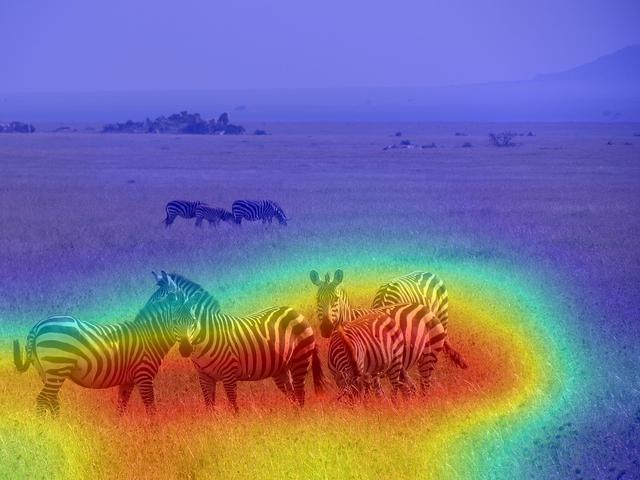} &
        \includegraphics[width=\imgsize, height=\imgsize]{imgs/100-loc-v2-ret/COCO_val2014_000000261796.jpg} &
        \includegraphics[width=\imgsize, height=\imgsize]{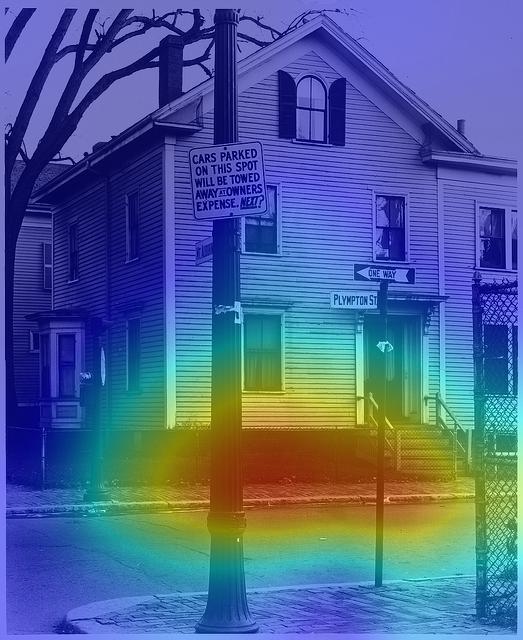} &
        \includegraphics[width=\imgsize, height=\imgsize]{imgs/100-loc-v2-ret/COCO_val2014_000000288685.jpg} &
        \includegraphics[width=\imgsize, height=\imgsize]{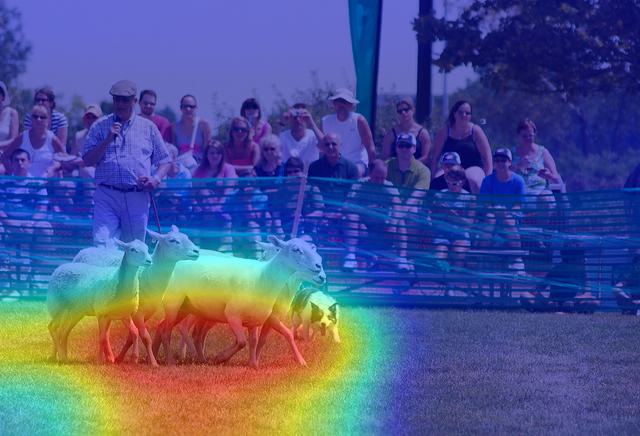} &
        \includegraphics[width=\imgsize, height=\imgsize]{imgs/100-loc-v2-ret/COCO_val2014_000000457559.jpg} &
        \includegraphics[width=\imgsize, height=\imgsize]{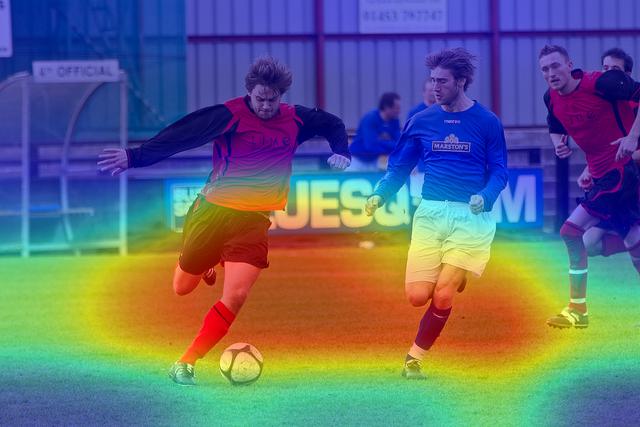} \\
        \multicolumn{2}{c|}{rank: 4 · is correct: \xmark} &
        \multicolumn{2}{c|}{rank: 4 · is correct: \cmark} &
        \multicolumn{2}{c|}{rank: 4 · is correct: \xmark} &
        \multicolumn{2}{c|}{rank: 4 · is correct: \xmark} &
        \multicolumn{2}{c }{rank: 4 · is correct: \cmark} \\
        \includegraphics[width=\imgsize, height=\imgsize]{imgs/100-loc-v2-ret/COCO_val2014_000000555412.jpg} &
        \includegraphics[width=\imgsize, height=\imgsize]{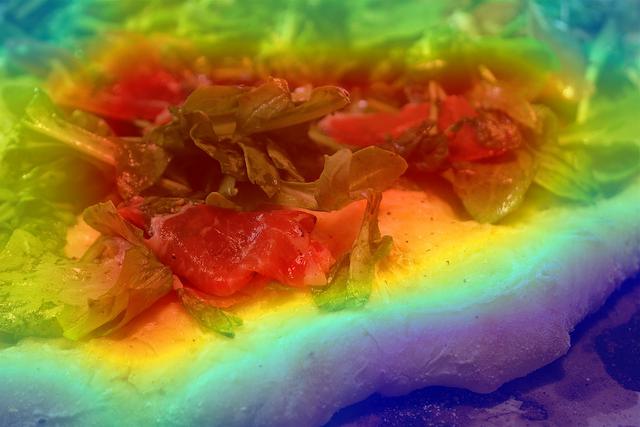} &
        \includegraphics[width=\imgsize, height=\imgsize]{imgs/100-loc-v2-ret/COCO_val2014_000000185409.jpg} &
        \includegraphics[width=\imgsize, height=\imgsize]{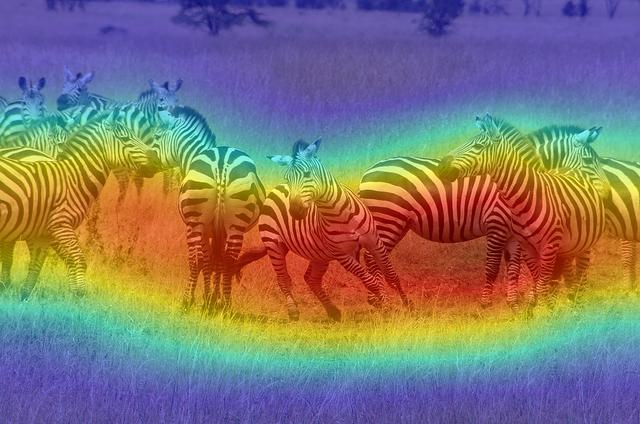} &
        \includegraphics[width=\imgsize, height=\imgsize]{imgs/100-loc-v2-ret/COCO_val2014_000000361142.jpg} &
        \includegraphics[width=\imgsize, height=\imgsize]{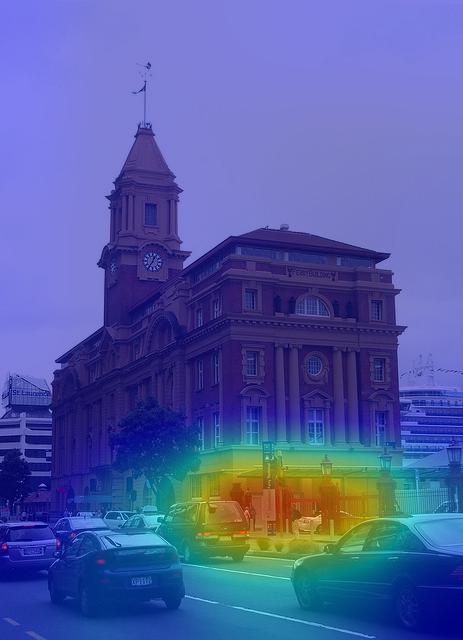} &
        \includegraphics[width=\imgsize, height=\imgsize]{imgs/100-loc-v2-ret/COCO_val2014_000000418281.jpg} &
        \includegraphics[width=\imgsize, height=\imgsize]{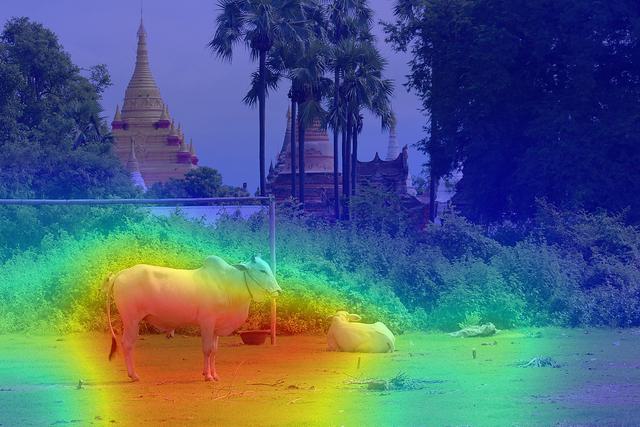} &
        \includegraphics[width=\imgsize, height=\imgsize]{imgs/100-loc-v2-ret/COCO_val2014_000000527528.jpg} &
        \includegraphics[width=\imgsize, height=\imgsize]{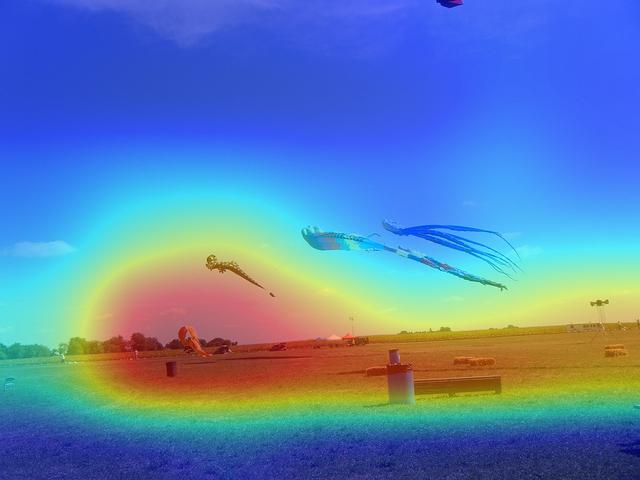} \\
        \multicolumn{2}{c|}{rank: 5 · is correct: \cmark} &
        \multicolumn{2}{c|}{rank: 5 · is correct: \cmark} &
        \multicolumn{2}{c|}{rank: 5 · is correct: \xmark} &
        \multicolumn{2}{c|}{rank: 5 · is correct: \xmark} &
        \multicolumn{2}{c }{rank: 5 · is correct: \cmark} \\
        \includegraphics[width=\imgsize, height=\imgsize]{imgs/100-loc-v2-ret/COCO_val2014_000000314182.jpg} &
        \includegraphics[width=\imgsize, height=\imgsize]{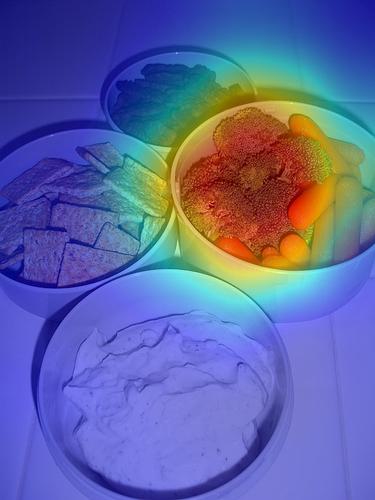} &
        \includegraphics[width=\imgsize, height=\imgsize]{imgs/100-loc-v2-ret/COCO_val2014_000000501243.jpg} &
        \includegraphics[width=\imgsize, height=\imgsize]{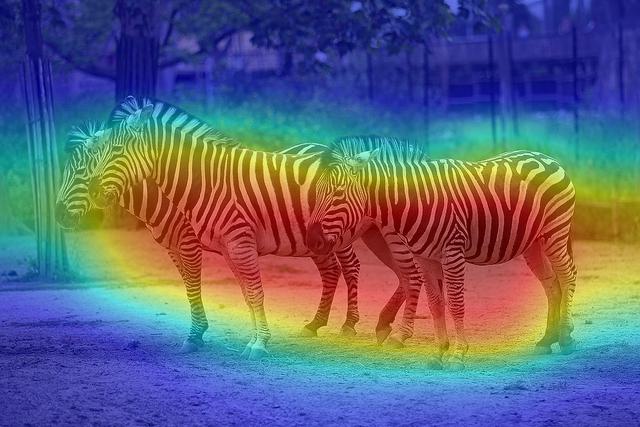} &
        \includegraphics[width=\imgsize, height=\imgsize]{imgs/100-loc-v2-ret/COCO_val2014_000000546823.jpg} &
        \includegraphics[width=\imgsize, height=\imgsize]{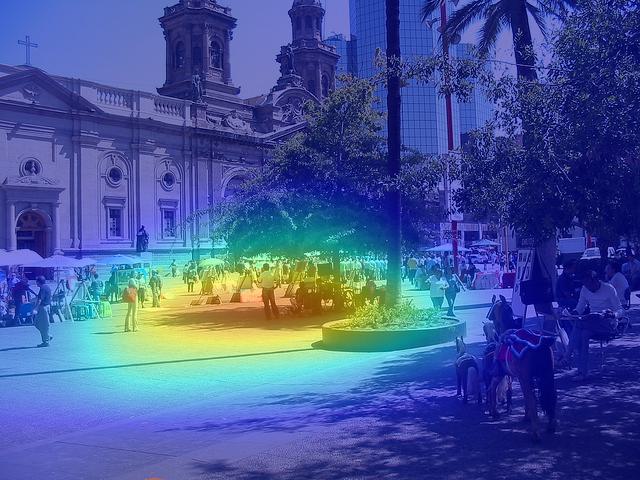} &
        \includegraphics[width=\imgsize, height=\imgsize]{imgs/100-loc-v2-ret/COCO_val2014_000000396205.jpg} &
        \includegraphics[width=\imgsize, height=\imgsize]{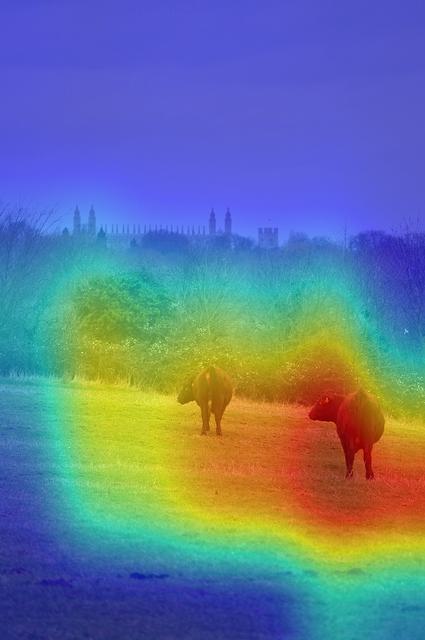} &
        \includegraphics[width=\imgsize, height=\imgsize]{imgs/100-loc-v2-ret/COCO_val2014_000000576654.jpg} &
        \includegraphics[width=\imgsize, height=\imgsize]{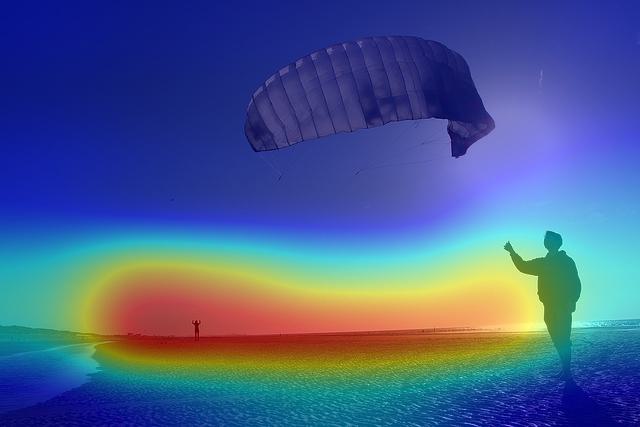} \\
	\end{tabular}
 	\caption{The top five ranked samples for audio query corresponding to each of the five concepts using the $K=100$ \smallmodel\ model. For each image, we show whether it is correct (if it contains the query concept) and the attention explanation (red indicating the input regions relevant for the given audio query).
    The IOU values give the quantitative localisation performance of the attention explanations:
    the intersection over union of the binarised attentions with the ground truth annotations averaged over all images that contain a given concept.
  }
	\label{tab:qualitative-results}
\end{figure*}

Where does \model\ focus in an image when given an audio query?
We make use of the implicit localisation capabilities provided by our model.
More precisely, since \model\ projects the image features to attention scores (based on the audio feature, \S\ref{subsec:model}),
we first reshape the 49-dimensional attention vector to a $7\times7$ matrix,
which we then resize %
with bilinear interpolation
to the original image size.
The resulting map reveals the importance of the input regions:
the closer the attention score is to 100, the more likely it is relevant to the input audio query.
Fig.~\ref{tab:qualitative-results} shows the top five retrieved images together with their attention maps for each of the five keywords.

These visualisations are useful for performing implicit localisation of the spoken concepts in the input images, but, perhaps more importantly, they allow us to better understand the model.
We observe that for ``broccoli'' the model selects the green food in an image; the ``fire hydrant'' queries are associated with urban scenes (streets and cars); and the keyword ``kite" is linked to the seaside or fields. %
While all these associations are useful proxies for identifying the spoken keywords, they also indicate that the model tends to learn correlations with the context or other spurious features.
A quantitative analysis in terms of the  intersection over union (IOU) with ground truth annotations supports these findings, as shown  at the top of Fig.~\ref{tab:qualitative-results}.
This effect is not surprising given that the model learns in a weakly-supervised manner (the training is done on full images with no explicit localisation information) and with noisy data (the mined samples are not always accurate).

With 
the analyses up to this point we can explain %
the discrepancies seen at the end of \S\ref{sec:english}.
Concretely, we conclude that the plateau seen in the retrieval scores (Table~\ref{tab:retrieval_task}) is due to the model associating contextual information or other spurious features with a few-shot class. %
E.g. associating anything green with \textit{broccoli}, urban areas or vertical pole-like objects with \textit{fire hydrant}, and people in fields with \textit{kite}.
This aids few-shot classification since the context or faulty features might help to distinguish the few-shot classes.
However, this hurts few-shot retrieval since the model will return green objects (\textit{field} or \textit{grass}) when prompted with ``broccoli''.

\subsection{Accuracy of the mining pairs}
\label{subsec:mining_pairs}

\begin{table}[tb]
	\caption{Precision %
 (\%) of the audio and image pairs mined from the $K$ few-shot examples in $\mathcal{S}$.}
    \captionsep
	\label{tab:pair_accuracy}
	\centering
	\footnotesize
	\setlength{\tabcolsep}{1em}
	\renewcommand{\arraystretch}{1.2}
	\begin{tabular}{rcc}
		\toprule
		$K$ &  Audio pairs &  Image pairs \\
		\midrule
		5 & 81.1 &  43.9\\
		10 & 83.0 & 47.5\\
		50 & 85.4 & 48.0 \\
		100 & 87.6 & 51.5\\
		\bottomrule
	\end{tabular}
\end{table}

In \S\ref{subsec:mining} we describe the method we use to mine training pairs from only the few ground truth word-image pairs in the support set $\mathcal{S}$: to find audio matches for each spoken keyword in $\mathcal{S}$ from a large collection of unlabelled speech, we use \qbert\ to extract possible word matches.
Similarly, we use AlexNet to find image matches for each image class in $\mathcal{S}$.
This means that the audio and image pairs we use for training are not 100\% correct. 
In order to evaluate the effect that the mined pairs have on \model's performance, we report the precision %
of these pairs in Table~\ref{tab:pair_accuracy}. 
For both the audio and image pairs, the precision increases as $K$ increases.

\subsection{Importance of the background data}
\label{subsec:background_images}

\begin{table}[!b]
	\caption{Few-shot retrieval scores (\%) obtained from \smallmodel\ %
    trained with and without negative background images.}
    \captionsep
	\label{tab:background_data_retrieval}
	\centering
	\footnotesize	
	\setlength{\tabcolsep}{0.1em}
	\renewcommand{\arraystretch}{1.2}
	\begin{tabular*}{\linewidth}{ p{12em}@{\extracolsep{\fill}}cccc }
		\toprule
		\multicolumn{1}{c}{Model} & 
		\multicolumn{4}{c}{$K$}\\
		\cline{2-5}
		 & 5 & 10 & 50 & 100 \\
		\midrule
		\smallmodel\ & \textbf{40.3$\pm$0.1} & \textbf{44.2$\pm$0.1} & \textbf{41.7$\pm$0.2} & \textbf{43.7$\pm$0.1}\\
          \smallmodel, no background data & 18.1$\pm$1.0 & 23.6$\pm$1.5 & 26.1$\pm$0.2 & 23.1$\pm$0.5\\
          \smallmodel, no background image negatives & 29.9$\pm$0.1 & 31.4$\pm$0.2 & 32.2$\pm$0.2 & 32.1$\pm$0.1\\
		\bottomrule
	\end{tabular*}
\end{table}

\begin{table}[!b]
    \caption{Few-shot classification accuracy (\%) obtained from \smallmodel\ %
    trained with and without negative background images.}
    \captionsep
    \label{tab:background_data_classification}
    \centering
    \footnotesize
    \setlength{\tabcolsep}{0.3em}
    \renewcommand{\arraystretch}{1.2}
    \begin{tabular*}{\linewidth}{l@{\extracolsep{\fill}}cccc}
        \toprule
        \multicolumn{1}{c}{Model} & \multicolumn{4}{c}{$K$}\\
        \cline{2-5}
         & 5 & 10 & 50 & 100 \\
         \midrule
         \smallmodel\ & 80.1 & 81.1 & 88.5 & 93.2\\
         \smallmodel, no background data & 65.1 & 64.7 & 75.3 & 77.5\\
         \smallmodel, no background image negatives & \textbf{88.0} & \textbf{90.1} & \textbf{94.8} & \textbf{95.3}\\
         \bottomrule
    \end{tabular*}
\end{table}

Ideally, the usage of background negative images should force the model to throw away more contextual information and focus more on learning the few-shot objects, since the background negatives might have similar contexts to some of the few-shot cases. 
But since we see that contextual information is not completely ignored in the above analyses, we do the analysis in Tables~\ref{tab:background_data_retrieval}~and~\ref{tab:background_data_classification}:
In the second line of both tables, we have a model which is not pretrained at all and also uses no image negatives during fine-tuning on the mined pairs (\S\ref{subsec:mining}), i.e.\ background data (\S\ref{sec:subsubsec_data}) is not used at all.
By comparing lines 1 and 2 of Table~\ref{tab:background_data_retrieval}, we see that retrieval scores drops substantially.
We can therefore conclude that
adding the background information does remove some contextual information, but not all of it.

Interestingly, we see that the classification scores also decrease (lines~1 and 2 of Table~\ref{tab:background_data_classification})  when we remove all background data. 
But this is mainly due to the effect of pretraining. 
To see this,
we look at the contribution of the background image negatives: we train a \model\ using the pretrained network but leave out the negative background images when fine-tuning. %
This model is listed in the third row of the two tables.
For classification (Table~\ref{tab:background_data_classification}), we see that this approach actually improves performance over \model\ (line 1). This makes sense since the model is now fine-tuned exclusively on (mined) few-shot classes. But it also illustrates that pretraining on background data is essential. For retrieval (Table~\ref{tab:background_data_retrieval}), we see an expected drop in performance comparing lines 1 and 3, because the latter model is not trained to distinguish between few-shot and non-few-shot classes.

\subsection{Adding more keywords}
\label{subsec:more_keywords}

Can we use our model to deal with more than five classes? 
We leave an exhaustive investigation of this question for future work, but present some initial experiments here.
Concretely, we do a test on 40 classes, which we manually select from the dataset.
When training \model\ on the 40 classes, performance on the same few-shot retrieval task for the original five classes drops marginally from 40.3\% to 37.1\%. 
This shows that, despite being trained on more classes, the model still retains most of its retrieval performance.  
Similarly, classification performance on the original five classes drops from 80.1\% to 74.3\%. 
When doing 40-way few-shot classification, we achieve a performance of 23.8\% (this is a much more difficult task than the five-way setting).

\begin{figure}[!b]
    \centering
    \includegraphics[width=0.9\linewidth]{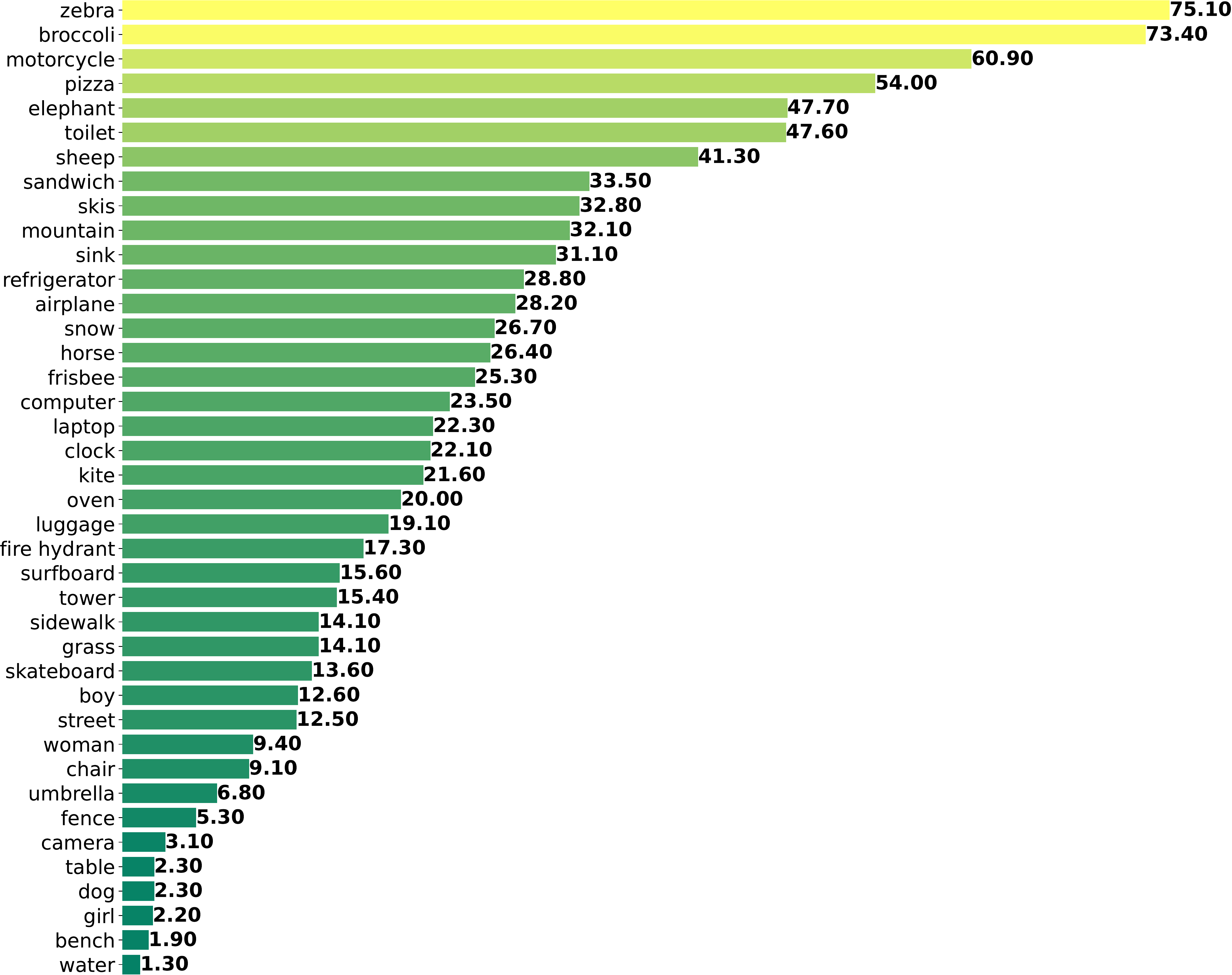}
    \caption{The per-keyword classification scores for the 40 few-shot classes.}
    \label{fig:per_keyword_40_classes}
\end{figure}

To see what happens on the per-keyword level when \model\ is trained and tested on 40 few-shot classes, Fig.~\ref{fig:per_keyword_40_classes} shows the individual scores.
With 40 classes, there is more room for error. %
We also see a larger distribution of few-shot classes that the model struggles to learn.
Future work will look into improving the mined image pairs specifically (since the audio pairs are more accurate). %

\section{Actual Low-Resource Few-Shot Speech-Image Experiments}

Our ultimate goal is to do multimodal few-shot word acquisition on actual low-resource languages.
To showcase our model's capabilities in this regard, we perform multimodal few-shot experiments on \yoruba, a low-resource language spoken in Nigeria by roughly 44 million people.
We apply \model\ (\S\ref{sec:model})
to the \yoruba\ Flickr Audio Caption Corpus (YFACC)~\cite{olaleye_yfacc_2023}.
Because the dataset is small, we consider only the case with five shots per class ($K = 5$).
We also train an English version of our model using a similarly sized dataset %
for comparison.

\subsection{Experimental setup}

\subsubsection{Data}
YFACC is an extension of the original English Flickr image-text captioning corpus~\cite{rashtchian_collecting_2010, hodosh_framing_2013}. %
Concretely, for each of the 8k Flickr images, one English text caption was translated into \yoruba\ and then recorded by a single speaker.
The YFACC paper~\cite{olaleye_yfacc_2023} looked specifically at a keyword spotting task for 67 \yoruba\ keywords (matching the English keywords from~\cite{kamper_semantic_2019}). 
The YFACC test set therefore contains 500 spoken captions with manual alignments where each caption contains at least one of their 67 keywords.
Since not all our original few-shot classes  occur in this set (\S\ref{sec:english_setup}), we choose five new classes from their 67 keywords. 
The \yoruba\ keywords with their English translations are given in Table~\ref{tab:new_flickr_keywords}.

\begin{table}[!b]
	\caption{The five new few-shot classes for the English and \yoruba\ Flickr experiments.}
 \captionsep
\label{tab:new_flickr_keywords}
\centering
\footnotesize
\renewcommand{\arraystretch}{1.2}	
\begin{tabular}{ cc}
	\toprule
	English & \yoruba\\
	\midrule
	boy & \boy\\
	dogs & \dogs\\
	grass & \grass\\
	rock & \rock\\
	water & \water\\
	\bottomrule
\end{tabular}
\end{table}

Since only the test set contains alignments for the keywords,
we have to sample the support set from the test set and use the remaining examples to sample episodes for the few-shot classification task.
For the support set images, we crop the images to only contain a single few-shot class since the new classes frequently co-occur within images (unlike the setting for the English experiments in \S\ref{sec:english}).
We stress that none of the examples in the support set occurs in the few-shot episodes' matching sets or queries.
We sample the episodes for the new Flickr keywords in the same manner as in \S\ref{subsubsec:few-shot_tasks_implementation}.

As a baseline, we also train an English system on very similar data. 
For this, we use the English Flickr Audio Captions Corpus (FACC)~\cite{harwath_deep_2015} which preceded and inspired YFACC.
We follow the same setup as for the \yoruba\ model, but replace the \yoruba\ 
utterances
from YFACC with the corresponding English 
utterances
from FACC.

\subsubsection{Models}

Using the %
sampled support set, we mine image and audio pairs in the same way as %
in \S\ref{subsec:mining_pairs}, with the only difference that we take the $n=100$ highest ranking examples per class 
because the dataset is smaller.
For mining audio pairs in \yoruba, we still use the QbERT approach, based on an English-trained HuBERT model (\S\ref{subsec:mining}), even though we are searching through unlabelled \yoruba\ audio.
I.e.\ we apply QbERT cross-lingually.
For the English model, we do not use weights pretrained on the MSCOCO background data as we did before, since this data might contain paired instances of our few-shot classes.
There is therefore no background speech-image pretraining.
However, we use the pretrained convolutional part of AlexNet, as well as the pretrained acoustic network of \cite{nortje_visually_2023} for initialisation (\S\ref{sec:models}).
The rest of the implementations remain the same as set out in \S\ref{sec:english_setup}.%

\subsubsection{Few-shot evaluation}

For evaluation, we only consider the few-shot classification task.
We report %
performance
over \num{1000} episodes,
each episode consisting of five spoken query words (one for each of the few-shot classes) and five images (also one for each of the few-shot classes) in the matching set.
In contrast to the English experiments in \S\ref{sec:english} and \S\ref{sec:english_ablation}, each matching image can potentially belong to more than one few-shot class. 
The reason for this is that the few-shot classes frequently co-occur within Flickr images.
To determine whether a particular image contains one of the few-shot classes, we use the ground truth text transcriptions available with the Flickr data (five text captions per image for English and one per image for \yoruba).
We mark a prediction as correct when the query audio word matches any of the words in a transcript.

\subsection{Experimental results}

\begin{table}[!b]
    \caption{%
        Few-shot classification accuracy (\%) on the Flickr data with English and three \yoruba\ versions of \smallmodel.
        The \yoruba\ \textsf{pt} model is a \yoruba\ model initialised with the weights of the pretrained SpokenCOCO English model from \S\ref{sec:models}.%
    }
    \captionsep
    \label{tab:yoruba}
    \newcommand{\cmark}{\color{black}\ding{51}}%
    \newcommand{\xmark}{\color{gray!50}\ding{55}}%
    \centering
    \footnotesize
    \begin{tabular*}{\linewidth}{l c c ccccc}
        \toprule
              & & Average  & \textit{boy} & \textit{dogs} & \textit{grass}      & \textit{rock} & \textit{water} \\
        Model & Mine & accuracy & {\tiny\boy} & {\tiny\dogs} & {\tiny\grass} & {\tiny\rock} & {\tiny\water} \\
        \midrule 
        Eng.              & \cmark & 59.4 & 56.9 & 28.9 & 64.3 & 78.4 & 68.3 \\
        \addlinespace
        Yor.             & \cmark & 36.3 & 18.8 & 23.2 & 35.4 & 49.8 & 54.5 \\
        Yor. \textsf{pt} & \cmark & 62.0 & 66.9 & 37.4 & 56.6 & 78.0 & 71.3 \\
        Yor. \textsf{pt} & \xmark & 29.4 & 47.1 & 30.7 & 29.7 & 21.1 & 18.2 \\
        \bottomrule
    \end{tabular*}
\end{table}

The multimodal few-shot classification results are given in Table~\ref{tab:yoruba}.
Line 2 represents the first time that multimodal few-shot word classification is performed on a real low-resource language, \yoruba.
We see that the performance of the \yoruba\ model on this task (36.3\%) is worse compared to the English baseline (59.4\%), trained and evaluated using a similar setup.

We didn't initialise the English baseline model here by pertaining it on the background SpokenCOCO data as we did before (\S\ref{sec:english_setup}) because this background data might contain paired English instances of the few-shot classes that we are using here.
However, it is a fair experiment to use the SpokenCOCO English data to initialise a \yoruba\ \model, i.e.\ we use the available resources from a well-resourced language to transfer knowledge to a low-resourced setting. 
The results for this approach are given in line~3 of Table~\ref{tab:yoruba}.
This \yoruba\ \model\ model outperforms both the English and \yoruba\ model without pretraining in terms of overall accuracy.
Per-keyword classification scores are also given in Table~\ref{tab:yoruba}.
The English-pretrained \yoruba\ model gives substantial improvements over the non-pretrained model across all five keyword classes, and also outperforms the English model on three of the few-shot classes.

\begin{figure}
    	\setlength{\tabcolsep}{2pt}
    	\newcommand{\cmark}{\color{indiagreen}\ding{51}}%
    	\newcommand{\xmark}{\color{red}\ding{55}}%
    	\centering
     \def\imgsize{1.5cm}
     \def\imgheight{0.75cm}
        \footnotesize
    	\begin{tabular}{ccccc}
    		\multicolumn{5}{c}{\color{gray} \textsc{English}} \\
    		\multicolumn{5}{c}{\color{gray} matching set \& annotations} \\
            \color{gray} \it boy &
            \color{gray} \it dogs &
            \color{gray} \it dogs, grass &
            \color{gray} \it rock &
            \color{gray} \it boy, water \\
            \includegraphics[width=\imgsize, height=\imgsize]{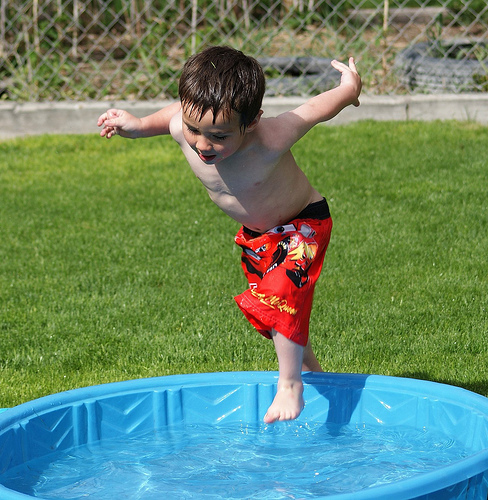} &
            \includegraphics[width=\imgsize, height=\imgsize]{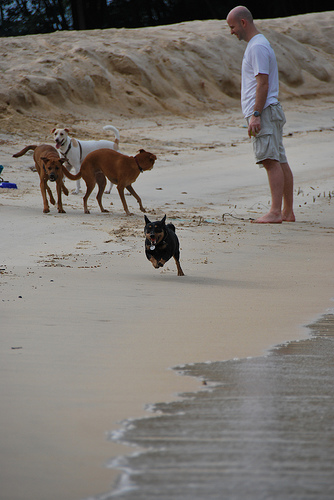} &
            \includegraphics[width=\imgsize, height=\imgsize]{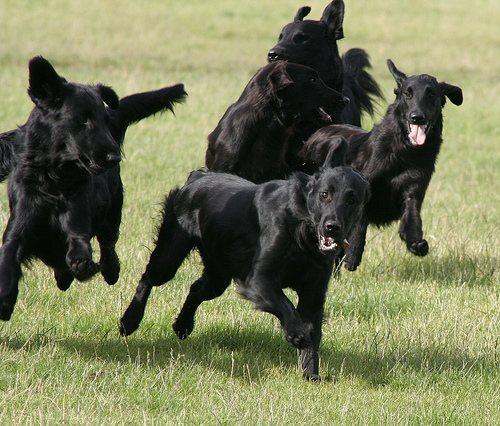} &
            \includegraphics[width=\imgsize, height=\imgsize]{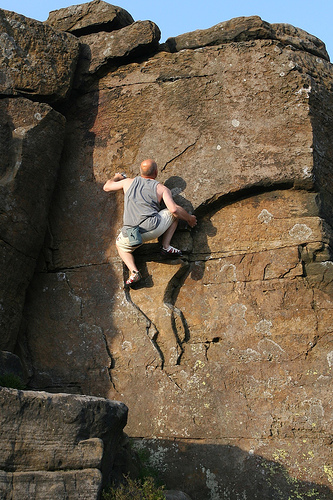} &
            \includegraphics[width=\imgsize, height=\imgsize]{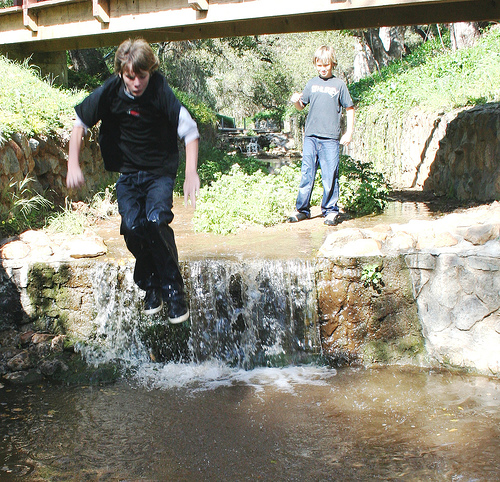} \\
    		\multicolumn{5}{c}{\color{gray} audio queries} \\
            \color{gray} \it boy &
            \color{gray} \it dogs &
            \color{gray} \it grass &
            \color{gray} \it rock &
            \color{gray} \it water \\
            \includegraphics[width=\imgsize, height=\imgheight]{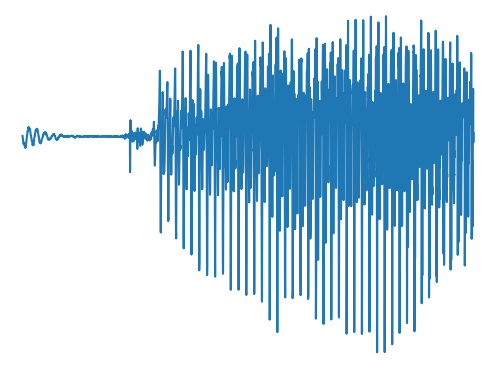} &
            \includegraphics[width=\imgsize, height=\imgheight]{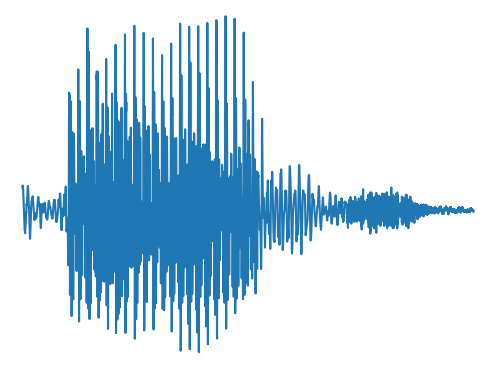} &
            \includegraphics[width=\imgsize, height=\imgheight]{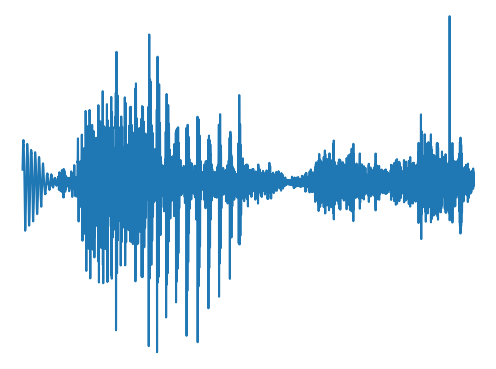} &
            \includegraphics[width=\imgsize, height=\imgheight]{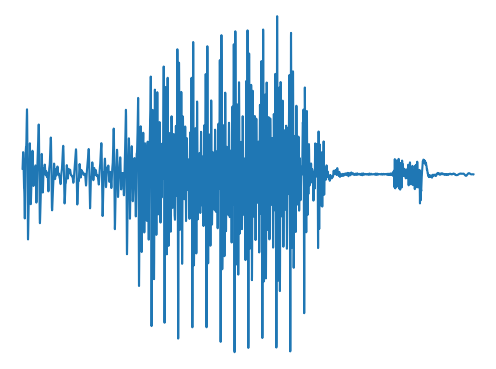} &
            \includegraphics[width=\imgsize, height=\imgheight]{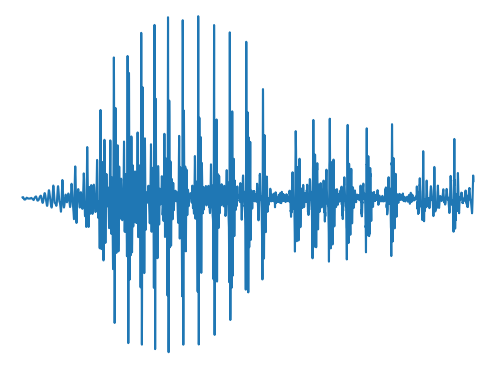} \\
    		\multicolumn{5}{c}{\color{gray} predicted images with attention, scores \& correctness} \\
            \includegraphics[width=\imgsize, height=\imgsize]{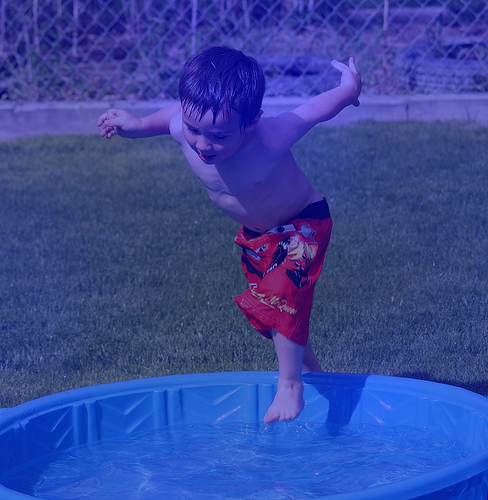} &
            \includegraphics[width=\imgsize, height=\imgsize]{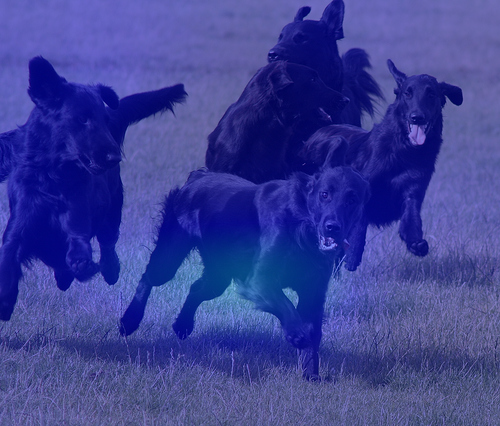} &
            \includegraphics[width=\imgsize, height=\imgsize]{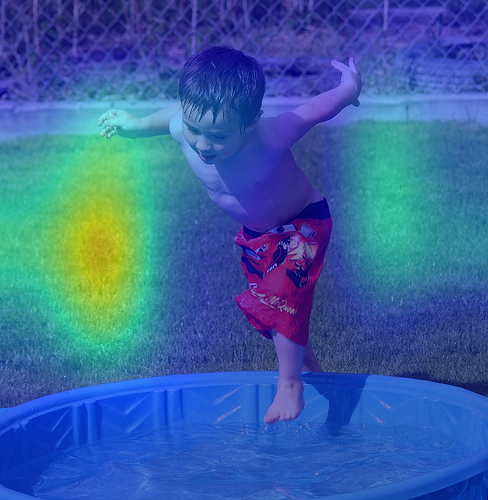} &
            \includegraphics[width=\imgsize, height=\imgsize]{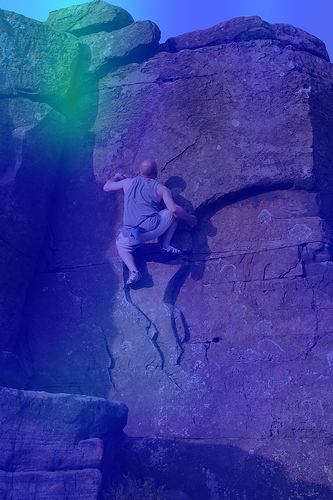} &
            \includegraphics[width=\imgsize, height=\imgsize]{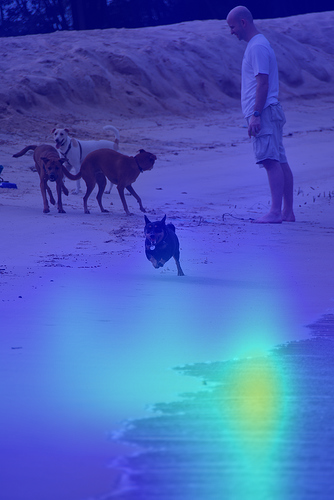} \\
             12.3 · \cmark &
             23.3 · \cmark &
             64.7 · \xmark &
             41.1 · \cmark &
             51.8 · \xmark \\
            \midrule
    		\multicolumn{5}{c}{\color{gray} \textsc{\yoruba}} \\
    		\multicolumn{5}{c}{\color{gray} matching set \& annotations} \\
            \color{gray} \it &
            \color{gray} \it &
            \color{gray} \it \dogs &
            \color{gray} \it &
            \color{gray} \it \\
            \color{gray} \it \boy &
            \color{gray} \it \dogs &
            \color{gray} \it \grass &
            \color{gray} \it \rock &
            \color{gray} \it \water \\
            \includegraphics[width=\imgsize, height=\imgsize]{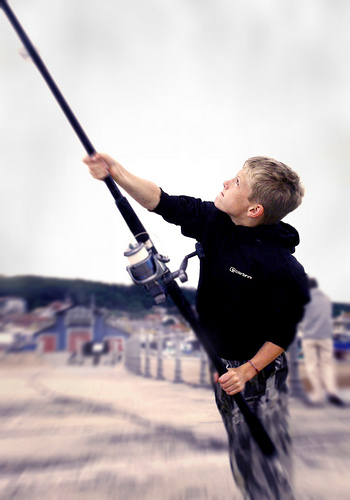} &
            \includegraphics[width=\imgsize, height=\imgsize]{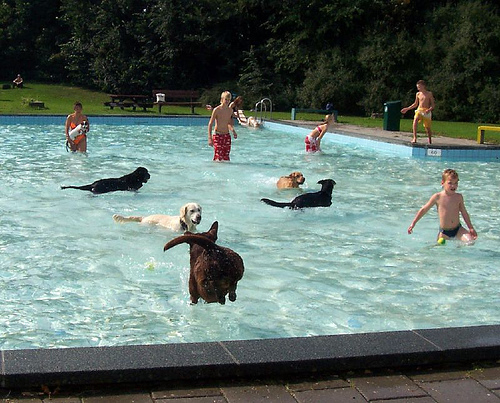} &
            \includegraphics[width=\imgsize, height=\imgsize]{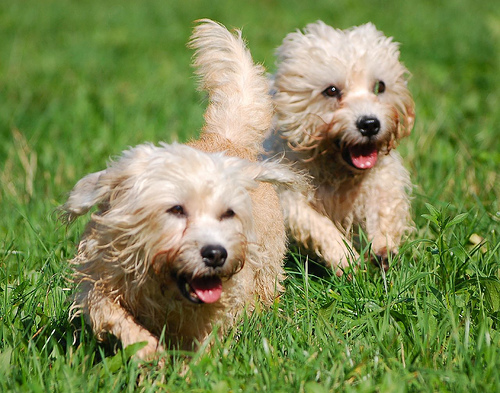} &
            \includegraphics[width=\imgsize, height=\imgsize]{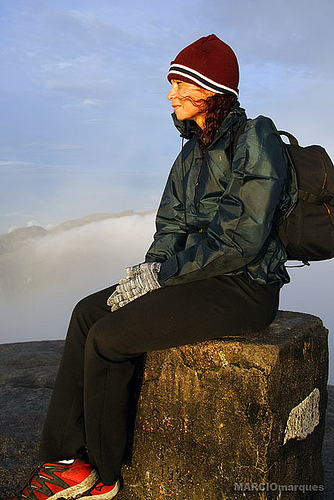} &
            \includegraphics[width=\imgsize, height=\imgsize]{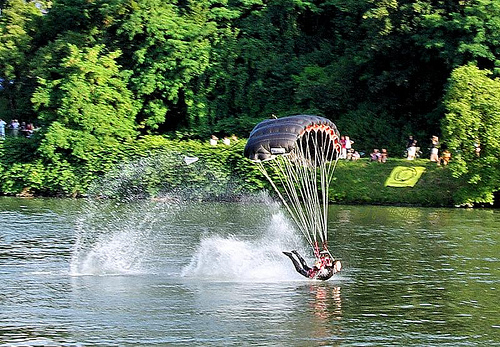} \\
            \multicolumn{5}{c}{\color{gray} audio queries} \\
            \color{gray} \it \boy &
            \color{gray} \it \dogs &
            \color{gray} \it \grass &
            \color{gray} \it \rock &
            \color{gray} \it \water \\
            \includegraphics[width=\imgsize, height=\imgheight]{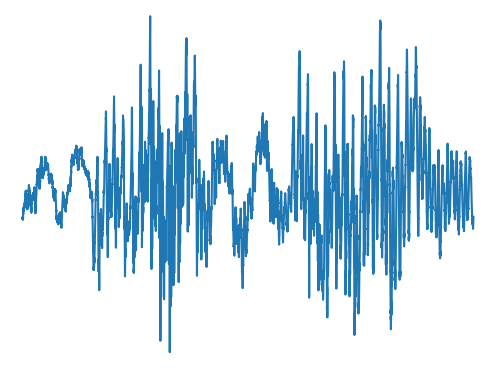} &
            \includegraphics[width=\imgsize, height=\imgheight]{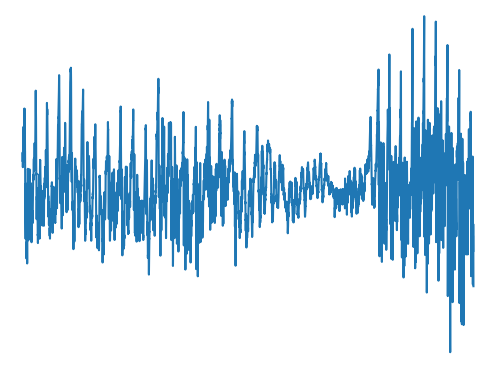} &
            \includegraphics[width=\imgsize, height=\imgheight]{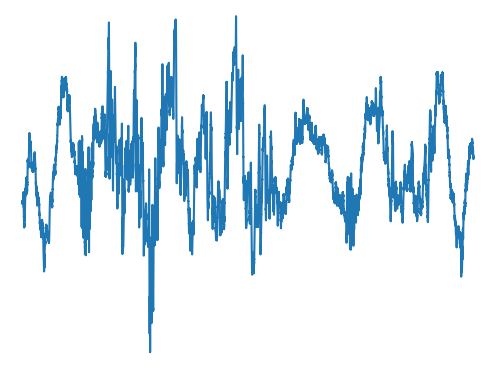} &
            \includegraphics[width=\imgsize, height=\imgheight]{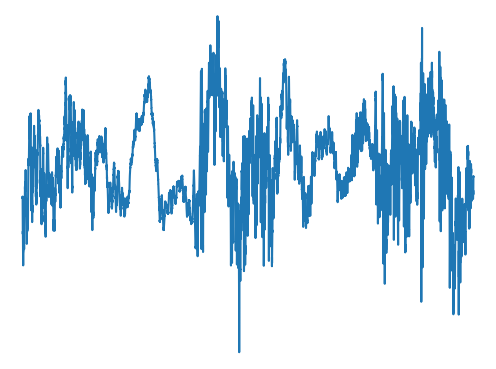} &
            \includegraphics[width=\imgsize, height=\imgheight]{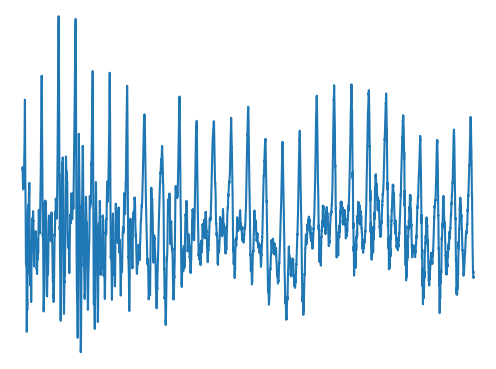} \\
    		\multicolumn{5}{c}{\color{gray} predicted images with attention, scores \& correctness} \\
            \includegraphics[width=\imgsize, height=\imgsize]{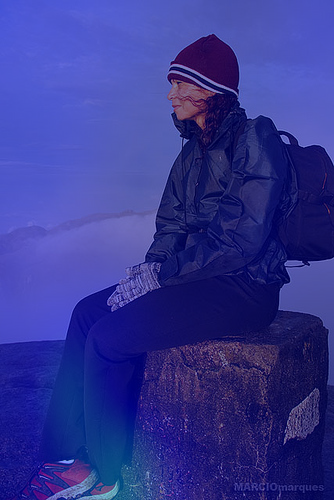} &
            \includegraphics[width=\imgsize, height=\imgsize]{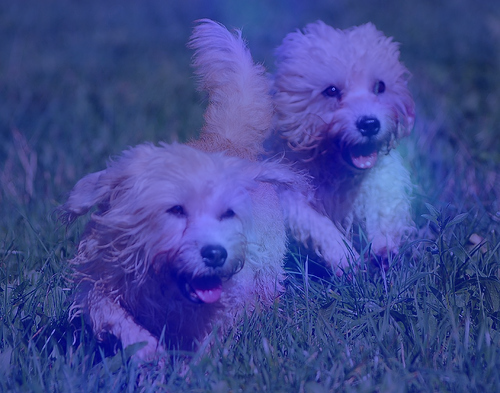} &
            \includegraphics[width=\imgsize, height=\imgsize]{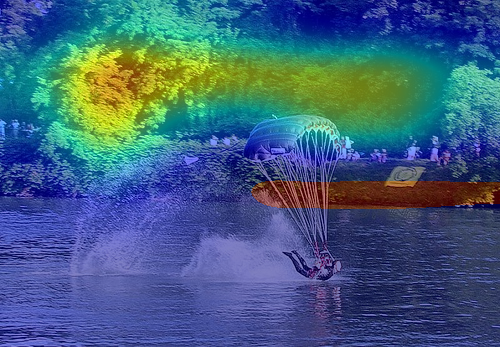} &
            \includegraphics[width=\imgsize, height=\imgsize]{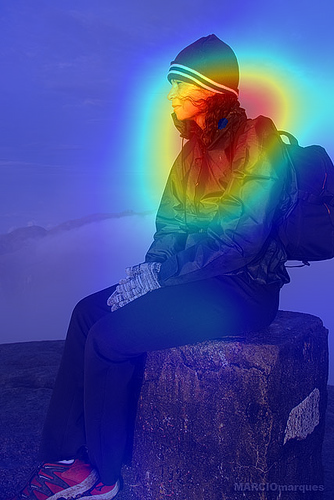} &
            \includegraphics[width=\imgsize, height=\imgsize]{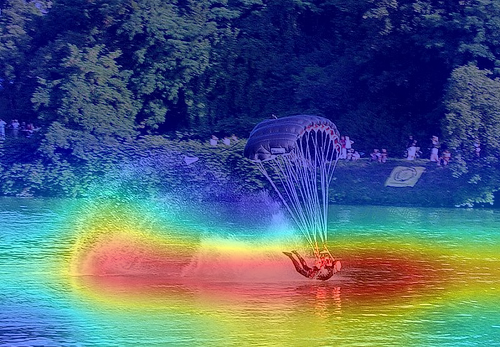} \\
             29.0 · \xmark &
             25.3 · \cmark &
             72.6 · \xmark &
            112.5 · \cmark &
            108.2 · \cmark \\
    	\end{tabular}
        \caption{Qualitative results showing the first few-shot classification episode on Flickr for English and \yoruba. For \yoruba\ we are showing results using the model with pretraining on background SpokenCOCO data. The labels are selected based on the text captions. We show attention weights overlaid on top of the images, with blue indicating the lowest scores and red the highest.}
    	\label{fig:flickr-qualitative}
\end{figure}

Fig. \ref{fig:flickr-qualitative} shows qualitative examples on the first episode for the English model and the \yoruba\ model with pretraining.
We observe that many of the predictions are accurate.
Some that are deemed to be wrong could in fact be considered correct. 
As a reminder, our evaluation is based on the captions associated with each image, and in some cases the captions might omit a keyword even though it is actually present in the image~\cite{berg_understanding_2012}.
E.g.\ ``grass'' is not mentioned in relation to the image with the boy jumping in the pool, but grass does occur in the image, and the model selects this image. 
So even though it is marked as a mistake in our evaluation protocol, it is actually correct.
The figure also shows that the pretrained \yoruba\ model seems more confident---the attention scores are higher for a given audio keyword---than the English one,
showing that the pretraining improved not only the overall performance but also the attention explanations.

Altogether, the experiments on \yoruba\ shows that \model\ can be applied to learn word-image classes from a few examples in a low-resource language when we also take advantage of a well-resourced language like English.
This is similar to the findings of~\cite{kamper_improved_2021, jacobs_acoustic_2021}.

\subsection{Mining performance across languages}
\label{subsec:mining-performance-flickr}

\begin{table}[!t]
    \caption{Precision scores (\%) of the mined English and \yoruba\ Flickr pairs, for image and audio mining.}
    \captionsep
    \label{tab:yoruba_vs_english_per_keyword_mining_precision}
    \centering
    \footnotesize
    \renewcommand{\arraystretch}{1.2}	
    \begin{tabular*}{\linewidth}{ l@{\extracolsep{\fill}}ccccc}
    	\toprule
            & \textit{boy}  & \textit{dogs}  & \textit{grass}  & \textit{rock} & \textit{water}\\
            Modality & {\scriptsize\boy} & {\scriptsize\dogs} & {\scriptsize\grass} & {\scriptsize\rock} & {\scriptsize\water}\\
    	\midrule
            Images & 38 & 46 & 62 & 19 & 80 \\
    	English audio  & 95 & 45 & 87 & 49 & 97 \\ 
    	\yoruba\ audio &  70 & 42 & 58 & 57 & 84 \\
        \bottomrule
    \end{tabular*}
\end{table}

The last question that remains is: what is the effect of using an English speech system, \qbert, to mine \yoruba\ audio pairs? 
Here we investigate the mining performance of the audio segments.
In Table \ref{tab:yoruba_vs_english_per_keyword_mining_precision} %
we show the precision of the mined audio pairs for the two languages and five keywords.
To get the precision scores, a mined audio segment is considered correct if the query keyword appears \emph{anywhere} in the caption (that is, we do not take the temporal alignments of the mined segment into consideration).%
\footnote{For English, the differences between this metric and its stricter variant that also checks that the temporal alignments of the mined segment agree with the spoken word are minute: less than 2\% absolute.}

We observe that for English three keywords (\textit{boy}, \textit{grass} and \textit{water}) obtain excellent results (around 90\% precision),
while the other two are performing more modestly (around 50\% precision).
To understand these results we find the most common mistakes for each of these three keywords; unsurprisingly, these mistakes are phonetically similar words:
``dogs'' is confused with ``dog'' (51 times out of the 100 mined samples) and ``dock'' (1), while ``rock'' is confused with ``dog'' (20) and ``rocky'' (13).
For \yoruba\ the performance across keywords has a more uniform spread and, while the top performance is lower than what we obtain on English, many of the keywords still obtain a reasonable precision.

Interestingly, the performance on the mined audio segments does not seem to necessarily correlate with the final performance.
E.g.\ while the mined audio pairs of \textit{rock} are among the worst,
its downstream performance is the best for both the English and the \yoruba\ pretrained models.
A similar observation can be made for \textit{boy}, which is accurately mined, but whose few-shot classification are poor compared to the other keywords.
One reason for this might be that the accuracy of the mined image pairs for \textit{boy} is low, as indicated in the first line of Table~\ref{tab:yoruba_vs_english_per_keyword_mining_precision} which shows the precision of the mined image pairs (which is the same for both languages). 

To quantify the effect of mining on classification performance,
we report results for the ``\yoruba\ \textsf{pt}, no mining'' model on each of the five keywords (Table~\ref{tab:yoruba}, row 4).
Comparing this to when mining is used (Table~\ref{tab:yoruba}, row 3),
we see that mining is beneficial even for keywords with modest mining performance.
This suggests that more positive samples, even noisy ones, are better than fewer clean ones.

\section{Conclusion}

Our goal was to do multimodal few-shot learning of natural images and spoken words. %
We proposed a novel few-shot pair mining method which we use in a new multimodal word-to-image attention model. %
For the %
scenario where the number of ``shots'' is small, our new model achieves higher few-shot retrieval scores than an existing model on a few-shot benchmark where an English query is used to retrieve images.
We also set a competitive baseline for natural visually grounded few-shot word classification using English data.

In further analyses, we showed that a few-shot model can be used to locate occurrences of an object in an image given a spoken query, and that many of the model's mistakes are due to associating contextual information with a few-shot class, e.g.\ \textit{fire-hydrant} often co-occurring with street views.

To showcase that our model can also be applied to a real low-resource language, we performed---for the first time---multimodal few-shot learning on a real low-resource language, \yoruba.
We showed that a multimodal \yoruba\ few-shot model can benefit substantially from being initialised on a more substantial amount of English speech-image data.

Future work will look into extending low-resource few-shot word classification to even more classes.  On both English and \yoruba\ data, our analysis also revealed that the image matching step in the mining scheme might, in particular, be limiting performance. Future work should therefore use recent advances from the vision community to improve image mining.

\bibliographystyle{IEEEtran}

\bibliography{mybib.bib}

\begin{thebibliography}{10}
\providecommand{\url}[1]{#1}
\csname url@samestyle\endcsname
\providecommand{\newblock}{\relax}
\providecommand{\bibinfo}[2]{#2}
\providecommand{\BIBentrySTDinterwordspacing}{\spaceskip=0pt\relax}
\providecommand{\BIBentryALTinterwordstretchfactor}{4}
\providecommand{\BIBentryALTinterwordspacing}{\spaceskip=\fontdimen2\font plus
\BIBentryALTinterwordstretchfactor\fontdimen3\font minus
  \fontdimen4\font\relax}
\providecommand{\BIBforeignlanguage}[2]{{%
\expandafter\ifx\csname l@#1\endcsname\relax
\typeout{** WARNING: IEEEtran.bst: No hyphenation pattern has been}%
\typeout{** loaded for the language `#1'. Using the pattern for}%
\typeout{** the default language instead.}%
\else
\language=\csname l@#1\endcsname
\fi
#2}}
\providecommand{\BIBdecl}{\relax}
\BIBdecl

\bibitem{besacier_automatic_2014}
L.~Besacier, E.~Barnard, A.~Karpov, and T.~Schultz, ``Automatic speech
  recognition for under-resourced languages: {A} survey,'' \emph{Speech
  Commun.}, 2014.

\bibitem{biederman_recognition-by-components:_1987}
I.~Biederman, ``Recognition-by-components: {A} theory of human image
  understanding.'' \emph{Psych. Review}, 1987.

\bibitem{miller_how_1987}
G.~Miller and P.~Gildea, ``How children learn words,'' \emph{SciAM}, 1987.

\bibitem{gomez_infant_2000}
R.~L. G{\'o}mez and L.~Gerken, ``Infant artificial language learning and
  language acquisition,'' \emph{TiCS}, 2000.

\bibitem{lake_one-shot_2014}
B.~M. Lake, C.-y. Lee, J.~R. Glass, and J.~B. Tenenbaum, ``One-shot learning of
  generative speech concepts,'' \emph{CogSci}, 2014.

\bibitem{rasanen_joint_2015}
O.~R{\"a}s{\"a}nen and H.~Rasilo, ``A joint model of word segmentation and
  meaning acquisition through cross-situational learning.'' \emph{Psych.
  Review}, 2015.

\bibitem{eloff_multimodal_2019}
R.~Eloff, H.~A. Engelbrecht, and H.~Kamper, ``Multimodal one-shot learning of
  speech and images,'' in \emph{Proc. {ICASSP}}, 2019.

\bibitem{nortje_unsupervised_2020}
L.~Nortje and H.~Kamper, ``Unsupervised vs. transfer learning for multimodal
  one-shot matching of speech and images,'' in \emph{Proc. {Interspeech}},
  2020.

\bibitem{nortje_direct_2021}
------, ``Direct multimodal few-shot learning of speech and images,'' in
  \emph{Proc. {Interspeech}}, 2021.

\bibitem{harwath_jointly_2018}
D.~Harwath, A.~Recasens, D.~Suris, G.~Chuang, A.~Torralba, and J.~Glass,
  ``Jointly discovering visual objects and spoken words from raw sensory
  input,'' in \emph{Proc. {ECCV}}, 2018.

\bibitem{kamper_semantic_2019-1}
H.~Kamper, A.~Anastassiou, and K.~Livescu, ``Semantic query-by-example speech
  search using visual grounding,'' in \emph{Proc. {ICASSP}}, 2019.

\bibitem{olaleye_attention-based_2021}
K.~Olaleye and H.~Kamper, ``Attention-based keyword localisation in speech
  using visual grounding,'' in \emph{Proc. {Interspeech}}, 2021.

\bibitem{chrupala_visually_2022}
G.~Chrupa{\l }a, ``Visually grounded models of spoken language: {A} survey of
  datasets, architectures and evaluation techniques,'' \emph{J. Artif. Intell.
  Res.}, 2022.

\bibitem{merkx_modelling_2022}
D.~Merkx, S.~Scholten, S.~L. Frank, M.~Ernestus, and O.~Scharenborg,
  ``Modelling human word learning and recognition using visually grounded
  speech,'' \emph{Cogn. Comput.}, 2022.

\bibitem{peng_syllable_2023}
P.~Peng, S.-W. Li, O.~R{\"a}s{\"a}nen, A.~Mohamed, and D.~Harwath, ``Syllable
  discovery and cross-lingual generalization in a visually grounded,
  self-supervised speech mode,'' in \emph{Proc. {Interspeech}}, 2023.

\bibitem{berry_m-speechclip_2023}
L.~Berry, Y.-J. Shih, H.-F. Wang, H.-J. Chang, H.-y. Lee, and D.~Harwath,
  ``M-{SpeechCLIP}: {Leveraging} large-scale, pre-trained models for
  multilingual speech to image retrieval,'' in \emph{Proc. {ICASSP}}, 2023.

\bibitem{miller_tyler_exploring_2022}
T.~Miller and D.~Harwath, ``Exploring few-shot fine-tuning strategies for
  models of visually grounded speech,'' in \emph{Proc. {Interspeech}}, 2022.

\bibitem{nortje_towards_2022}
L.~Nortje and H.~Kamper, ``Towards visually prompted keyword localisation for
  zero-resource spoken languages,'' in \emph{Proc. {SLT}}, 2022.

\bibitem{hsu_text-free_2021}
W.-N. Hsu, D.~Harwath, C.~Song, and J.~Glass, ``Text-free image-to-speech
  synthesis using learned segmental units,'' in \emph{Proc {ACL}}, 2021.

\bibitem{olaleye_yfacc_2023}
K.~Olaleye, D.~Oneata, and H.~Kamper, ``{YFACC}: {A} {Yor{\`u}b{\'a}}
  speech-image dataset for cross-lingual keyword localisation through visual
  grounding,'' in \emph{Proc. {SLT}}, 2023.

\bibitem{nortje_visually_2023}
L.~Nortje, B.~van Niekerk, and H.~Kamper, ``Visually grounded few-shot word
  acquisition with fewer shots,'' in \emph{Proc. {Interspeech}}, 2023.

\bibitem{he_deep_2016}
K.~He, X.~Zhang, S.~Ren, and J.~Sun, ``Deep residual learning for image
  recognition,'' in \emph{Proc. {CVPR}}, 2016.

\bibitem{krizhevsky_imagenet_2017}
A.~Krizhevsky, I.~Sutskever, and G.~E. Hinton, ``{ImageNet} classification with
  deep convolutional neural networks,'' \emph{ACM}, 2017.

\bibitem{kamper_truly_2019}
H.~Kamper, ``Truly unsupervised acoustic word embeddings using weak top-down
  constraints in encoder-decoder models,'' in \emph{Proc. {ICASSP}}, 2019.

\bibitem{chung_unsupervised_2016}
Y.-A. Chung, C.-C. Wu, C.-H. Shen, and H.-y. Lee,
  ``\BIBforeignlanguage{en}{Unsupervised learning of audio segment
  representations using sequence-to-sequence recurrent neural networks},'' in
  \emph{\BIBforeignlanguage{en}{Proc. {Interspeech}}}, 2016.

\bibitem{wang_segmental_2018}
Y.-H. Wang, H.-y. Lee, and L.-s. Lee, ``Segmental audio {Word2Vec}:
  {Representing} utterances as sequences of vectors with applications in spoken
  term detection,'' in \emph{Proc. {ICASSP}}, 2018.

\bibitem{holzenberger_learning_2018}
N.~Holzenberger, M.~Du, J.~Karadayi, R.~Riad, and E.~Dupoux, ``Learning word
  embeddings: {Unsupervised} methods for fixed-size representations of
  variable-length speech segments,'' in \emph{Proc. {Interspeech}}, 2018.

\bibitem{hsu_hubert_2021}
W.-N. Hsu, B.~Bolte, Y.-H.~H. Tsai, K.~Lakhotia, R.~Salakhutdinov, and
  A.~Mohamed, ``{HuBERT}: {Self}-supervised speech representation learning by
  masked prediction of hidden units,'' \emph{ACM}, 2021.

\bibitem{kamper_towards_2021}
H.~Kamper and B.~van Niekerk, ``Towards unsupervised phone and word
  segmentation using self-supervised vector-quantized neural networks,'' in
  \emph{Proc. {Interspeech}}, 2021.

\bibitem{needleman_general_1970}
S.~B. Needleman and C.~D. Wunsch, ``A general method applicable to the search
  for similarities in the amino acid sequence of two proteins,'' \emph{J. Mol.
  Biol.}, 1970.

\bibitem{lin_microsoft_2014}
T.-Y. Lin, M.~Maire, S.~Belongie, L.~Bourdev, R.~Girshick, J.~Hays, P.~Perona,
  D.~Ramanan, C.~L. Zitnick, and P.~Doll{\'a}r, ``Microsoft {COCO}: {Common}
  objects in context,'' in \emph{Proc. {ECCV}}, 2014.

\bibitem{mcauliffe_montreal_2017}
M.~McAuliffe, M.~Socolof, S.~Mihuc, M.~Wagner, and M.~Sonderegger, ``Montreal
  forced aligner: {Trainable} text-speech alignment using {Kaldi},'' in
  \emph{Proc. {Interspeech}}, 2017.

\bibitem{deng_imagenet_2009}
J.~Deng, W.~Dong, R.~Socher, L.-J. Li, K.~Li, and L.~Fei-Fei, ``{ImageNet}: {A}
  large-scale hierarchical image database,'' in \emph{Proc. {CVPR}}, 2009.

\bibitem{panayotov_librispeech_2015}
V.~Panayotov, G.~Chen, D.~Povey, and S.~Khudanpur, ``Librispeech: {An} {ASR}
  corpus based on public domain audio books,'' in \emph{Proc. {ICASSP}}, 2015.

\bibitem{harwath_vision_2018}
D.~Harwath, G.~Chuang, and J.~Glass, ``Vision as an interlingua: {Learning}
  multilingual semantic embeddings of untranscribed speech,'' in \emph{Proc.
  {ICASSP}}, 2018.

\bibitem{van_niekerk_vector-quantized_2020}
B.~van Niekerk, L.~Nortje, and H.~Kamper, ``Vector-quantized neural networks
  for acoustic unit discovery in the {ZeroSpeech} 2020 challenge,'' in
  \emph{Proc. {Interspeech}}, 2020.

\bibitem{harwath_unsupervised_2016}
D.~Harwath, A.~Torralba, and J.~Glass, ``Unsupervised learning of spoken
  language with visual context,'' in \emph{Proc. {NeurIPS}}, 2016.

\bibitem{kingma_adam_2015}
D.~Kingma and J.~Ba, ``Adam: {A} method for stochastic optimization,'' in
  \emph{Proc. {ICLR}}, 2015.

\bibitem{rashtchian_collecting_2010}
C.~Rashtchian, P.~Young, M.~Hodosh, and J.~Hockenmaier, ``Collecting {Image}
  {Annotations} {Using} {Amazon}'s {Mechanical} {Turk},'' in \emph{{NAACL}
  {HLT} {Workshop}}, 2010.

\bibitem{hodosh_framing_2013}
M.~Hodosh, P.~Young, and J.~Hockenmaier, ``Framing image description as a
  ranking task: {Data}, models and evaluation metrics,'' \emph{JAIR}, 2013.

\bibitem{kamper_semantic_2019}
H.~Kamper, G.~Shakhnarovich, and K.~Livescu, ``Semantic speech retrieval with a
  visually grounded model of untranscribed speech,'' \emph{IEEE/ACM Trans.
  Audio Speech Lang. Process.}, 2019.

\bibitem{harwath_deep_2015}
D.~Harwath and J.~Glass, ``Deep multimodal semantic embeddings for speech and
  images,'' in \emph{Proc. {ASRU}}, 2015.

\bibitem{berg_understanding_2012}
A.~C. Berg, T.~L. Berg, H.~Daum{\'e}, J.~Dodge, A.~Goyal, X.~Han, A.~Mensch,
  M.~Mitchell, A.~Sood, K.~Stratos, and K.~Yamaguchi, ``Understanding and
  predicting importance in images,'' in \emph{Proc. {CVPR}}, 2012.

\bibitem{kamper_improved_2021}
H.~Kamper, Y.~Matusevych, and S.~Goldwater, ``Improved acoustic word embeddings
  for zero-resource languages using multilingual transfer,'' \emph{TASLP},
  2021.

\bibitem{jacobs_acoustic_2021}
C.~Jacobs, Y.~Matusevych, and H.~Kamper, ``Acoustic word embeddings for
  zero-resource languages using self-supervised contrastive learning and
  multilingual adaptation,'' in \emph{Proc. {SLT}}, 2021.

\end{thebibliography}

\end{document}